\documentclass[12pt,draftclsnofoot,onecolumn]{IEEEtran}
\usepackage{graphicx}
\usepackage{color}
\usepackage{placeins}
\usepackage{float}
\usepackage{tabularx,colortbl}
\usepackage{amssymb}
\usepackage{amsthm}
\usepackage{cite}
\usepackage{amsmath}

\usepackage{caption2}

\usepackage{algorithm}
\usepackage{algpseudocode}
\theoremstyle{plain}
\newtheorem{thm}{Theorem}%%[section]

\theoremstyle{plain}
\newtheorem{rem}{Remark}
\newtheorem{cor}{Corollary}

\IEEEoverridecommandlockouts

\begin{document}

%----------------------------title&author&thanks----------------------------
%\title{Trajectory Design for Energy-Harvesting and Hardware Distorted UAV in CF Massive MIMO Systems}
%\title{Cell-Free Massive MIMO-OFDM for High-Speed Train Communications}
\title{Asynchronous Cell-Free Massive MIMO With Rate-Splitting}

\author{Jiakang Zheng, Jiayi Zhang,~\IEEEmembership{Senior Member,~IEEE}, Julian Cheng,~\IEEEmembership{Fellow,~IEEE}, Victor C. M. Leung,~\IEEEmembership{Life Fellow,~IEEE},\\ Derrick Wing Kwan Ng,~\IEEEmembership{Fellow,~IEEE}, and Bo Ai,~\IEEEmembership{Fellow,~IEEE}

%\thanks{
%This work was supported in part by National Key
%R\&D Program of China under Grant 2020YFB1807201, in part by National
%Natural Science Foundation of China under Grants 61971027, and 61961130391, in part by Beijing Natural Science Foundation under
%Grant L202013, in part by Natural Science Foundation of Jiangsu Province, Major Project under
%Grant BK20212002, in part by the Royal Society Newton Advanced Fellowship under Grant NA191006, in part
%by Frontiers Science Center for Smart High-speed Railway System.}

%\thanks{J. Zheng, and J. Zhang, are with the School of Electronic and Information Engineering, Beijing Jiaotong University, Beijing 100044, China (e-mail: \{jiakangzheng, jiayizhang\}@bjtu.edu.cn).}
%\thanks{J. Cheng is with the School of Engineering, The University of British Columbia, Kelowna, BC V1V 1V7, Canada (e-mail: julian.cheng@ubc.ca).}
%\thanks{V. C. M. Leung is with the College of Computer Science and Software Engineering, Shenzhen University, Shenzhen 518060, China, and also with the Department of Electrical and Computer Engineering, The University of British Columbia, Vancouver, BC V6T 1Z4, Canada (e-mail: vleung@ece.ubc.ca).}
%\thanks{D. W. K. Ng is with the School of Electrical Engineering and Telecommunications, University of New South Wales, Sydney, NSW 2052, Australia (e-mail: w.k.ng@unsw.edu.au).}
%\thanks{B. Ai is with the State Key Laboratory of Rail Traffic Control and Safety, Beijing Jiaotong University, Beijing 100044, China (e-mail: boai@bjtu.edu.cn).}

\thanks{J. Zheng, J. Zhang, and B. Ai are with Beijing Jiaotong University, China; J. Cheng is with the University of British Columbia, Canada; V. C. M. Leung is with Shenzhen University, China, and also with the University of British Columbia, Canada; D. W. K. Ng is with the University of New South Wales, Australia.}
}

\maketitle
\vspace{-1.8cm}
%----------------------------abstract----------------------------
\begin{abstract}
%Cell-free (CF) massive multiple-input multiple-output (MIMO) is considered as a promising technology for achieving the ultimate performance limit.
In practical cell-free (CF) massive multiple-input multiple-output (MIMO) networks with distributed and low-cost access points, the asynchronous arrival of signals at the user equipments increases multi-user interference that degrades the system performance.
Meanwhile, rate-splitting (RS), exploiting the transmission of both common and private messages, has demonstrated to offer considerable spectral efficiency (SE) improvements and its robustness against channel state information (CSI) imperfection.
The signal performance of a CF massive MIMO system is first analyzed for asynchronous reception capturing the joint effects of propagation delays and oscillator phases of transceivers.
Taking into account the imperfect CSI caused by asynchronous phases and pilot contamination, we derive novel and closed-form downlink SE expressions for characterizing the performance of both the RS-assisted and conventional non-RS-based systems adopting coherent and non-coherent data transmission schemes, respectively.
%Moreover, by exploiting the bisection method to solve the optimization problem that max-min per-user SE of common message, we propose a design for robust precoding of the common message to maximize the sum SE of the considered system.
Moreover, we formulate the design of robust precoding for the common messages as an optimization problem that maximizes the minimum individual SE of the common message. To address the non-convexity of the design problem, a bisection method is proposed to solve the problem optimally.
Simulation results show that asynchronous reception indeed destroys both the orthogonality of the pilots and the coherent data transmission resulting in poor system performance. Besides, thanks to the uniform coverage properties of CF massive MIMO systems, RS with a simple low-complexity precoding for the common message obtained by the equal ratio sum of the private precoding is able to achieve substantial downlink sum SE gains, while the application of robust precoding to the common message is shown to be useful in some extreme cases, e.g., serious oscillator mismatch and unknown delay phase.
\end{abstract}
\vspace{-0.4cm}

\IEEEpeerreviewmaketitle

\section{Introduction}

Cell-free (CF) massive multiple-input multiple-output (MIMO) has been envisioned as a disruptive technology to provide uniform spectral efficiency (SE) improvement and ubiquitous connectivity for the sixth-generation (6G) wireless communication networks \cite{9205230}. The key idea of CF massive MIMO systems is that through spatial multiplexing on the same time-frequency resources, a large number of geographically distributed access points (APs) connected to a central processing unit (CPU) are deployed to serve the user equipments (UEs) coherently \cite{Ngo2017Cell,9309348}.
Thanks to the law of large numbers, two phenomena characterizing the propagation environment of CF massive MIMO systems are channel hardening and favorable propagation, which facilitate simple precoding design and interference management \cite{8379438,9585108}. Additionally, the user-centric paradigm prevents the UE from perceiving the cell boundary, greatly simplifying its actual implementation \cite{bjornson2020scalable}.
In addition, the rich macro-diversity gain of CF massive MIMO systems brought by the joint transmission and reception has drawn much academic research interest in this topic \cite{zhang2020prospective}. For instance, it was shown that CF massive MIMO systems can outperform small-cell systems in terms of 95\%-likely per-user SE due to the joint interference cancellation capability of the former \cite{bjornson2019making}.
In addition, compared with traditional cellular networks, CF massive MIMO systems-based joint signal processing can effectively alleviate the influence of non-ideal practical factors, such as channel aging, hardware impairments, etc. \cite{9322468,zheng2021uav,9743355}, thanks to the large number of spatial degrees of freedom.
Moreover, it was revealed in \cite{demir2020joint} that the joint power control in CF massive MIMO systems is able to enhance the wireless power transfer efficiency compared with its counterpart in conventional systems. Besides, it was proved that the error probabilities of both the centralized and distributed joint activity detection in CF massive MIMO systems approach zero when the number of APs is sufficiently large \cite{9585108}. Furthermore, numerous joint optimization algorithms were proposed for CF massive MIMO systems to enhance the practicality of the system, including achieving the energy-efficient load balancing \cite{9136914}, reducing the complexity of channel estimation and decoding \cite{9507360}, and establishing a scalable framework by AP scheduling \cite{9309348}. Therefore, joint coherent processing is an important and fundamental topic for the practical implementation of CF massive MIMO systems.

Despite the fruitful research in the literature, most current works on CF massive MIMO assume the availability of perfect synchronization to ensure efficient joint coherent processing. Yet, this assumption is impractical for communication networks adopting a distributed architecture \cite{9650567}. One reason is that the geographically distributed APs cause inevitable signal arrival time differences from the UEs \cite{8676341}. In particular, the incurred delay phases impair the received signals creating a challenge for distributed massive MIMO systems to achieve coherent transmission. Another reason is that the transmitter and receiver hardware are not perfect and identical such that the imperfection may cause different random phase shifts on the channels \cite{6902790,9502552}. Besides, these oscillator phases introduce a multiplicative factor to the channel and this factor drifts gradually over time at each channel use following the Wiener process \cite{bjornson2015massive}. Unfortunately,
%the asynchronous reception issues in distributed massive MIMO mainly arise from two major factors: the delay phase caused by the distributed architecture and the oscillator phase caused by the low-cost hardware.
the impairments hinder the acquisition of accurate channel state information (CSI) and lead to severe multi-user interference. These jeopardize the achievable signal-to-noise ratio (SNR) and remain a major obstacle for the practical deployment of distributed massive MIMO systems \cite{9531352}.
In fact, accurate synchronization (e.g., phase, frequency) has always been an important and interesting research subject in the evolution of MIMO systems. For example, a profound study on the optimal estimator-detector receivers for the joint detection and synchronization was initiated for conventional digital communication \cite{meyr1998digital}. More importantly, the results indicated that effective coherent multiuser joint processing is only possible when a sufficient level of relative timing and phase synchronization accuracy is achieved at the APs \cite{6760595}.
As such, three effective approaches, from centralized to distributed implementations, were proposed to achieve phase synchronization in coherent MIMO radar systems \cite{5958644}. Besides, a high-accuracy frequency synchronization technique was proposed for the uplink of multiuser orthogonal-frequency-division multiplexing-based massive MIMO systems \cite{8691589}. However, the computational complexity of these algorithms increase significantly with the communication network size. In addition, although inserting a time-frequency interval to a resource block is an effective method to resolve the synchronization issues, it requires significant system resources for signaling overhead that may in turn reduce the overall system rate performance \cite{7913686}. Moreover, a recent performance analysis of CF massive MIMO systems for revealing the impacts of asynchronous reception was investigated \cite{9531352,8676341}.
Yet, an effective multi-user interference management for the case of imperfect CSI incurred by asynchronous CF massive MIMO systems has not been reported in the literature.

Recently, rate-splitting (RS) has been developed to harness multiuser interference particularly in the existence of imperfect CSI \cite{7470942}, showing its rich potential in addressing the negative impacts caused by asynchronous signal receptions in CF massive MIMO systems.
Indeed, the RS strategy divides the message for each user into two parts: a common message to be decoded by all the UEs and a private message to be decoded by the corresponding UE only \cite{mao2018rate}, respectively. Then, all the common parts are combined into a super common message and superposition coding is used to transmit these message streams simultaneously. At the UE, the common message is decoded first, with all the private messages treated as noise. Then, the disired private message is decoded after the common message has been removed via successive interference cancellation (SIC) \cite{7513415}.
The ability of RS to generalize two extreme existing approaches, i.e., treating interference as noise and interference for decoding, is what makes this strategy attractive for practical implementation \cite{7434643}.
For instance, the RS-based beamforming scheme performs strictly better than that based on non-orthogonal multiple access (NOMA) in both partially and fully loaded systems \cite{8019852}.
Besides, compared to conventional linear precoding approaches, the RS-assisted non-orthogonal unicast and multicast transmission techniques are more spectrally and energy efficient in a wide range of user deployment and network load scenarios \cite{8846706}. In addition to the system performance gains, relaxed CSI quality requirements and enhanced achievable rate regions are two other benefits of RS \cite{7555358}.
Moreover, it was proved that the RS strategy is a robust and handy alternative to conventional methods for mitigating the detrimental effects of hardware impairments \cite{7892949}, mobility \cite{9491092}, and limited feedback \cite{7434643} in massive MIMO systems.
In fact, the advantages of RS in massive MIMO can be well retained in CF massive MIMO systems, e.g., mitigating the pilot contamination \cite{9942944}.
To the best of the authors' knowledge, the research of RS in addressing the issues in CF massive MIMO systems is still incomplete and not thorough.
Therefore, thorough analysis and insights are required to unleash the potential of RS in enhancing the performance the CF massive MIMO system.

Motivated by the aforementioned observations, we first analyze the downlink performance of CF massive MIMO systems with asynchronous reception, including both the delay and oscillator phases. To alleviate the multiuser interference caused by imperfect CSI, we implement RS based on the transmission of common and private messages. Finally, we propose a robust precoding for the common message to further reduce the impact of asynchronous reception on CF massive MIMO systems. The specific contributions of this work are listed as follows:

\begin{itemize}
  \item Taking into account the imperfect CSI caused by asynchronous phase and pilot contamination, we first derive novel and closed-form downlink SE expressions for coherent and non-coherent data transmission schemes, respectively. Our results show that the existence of asynchronous reception destroys the orthogonality of pilots and the coherent transmission of data, leading to an unsatisfactory SE performance of CF massive MIMO systems.
  \item To compensate the system performance loss, we apply the RS strategy to CF massive MIMO systems by splitting the messages into common and private parts. We derive the closed-from sum SE expression for the RS-assisted CF massive MIMO system with asynchronous reception adopting an arbitrary linear precoding of the common messages. We also develop an efficient method to determine the optimal power allocating to the two messages. It is discovered that RS can considerably enhance the performance of CF massive MIMO systems. However, RS is negatively impacted by asynchronous phase, particularly the oscillator phase of the UE and delay phase.
  \item We then propose a design of robust precoding for the common messages to enhance the performance of RS-assisted CF massive MIMO systems by using the bisection method to solve the optimization problem that maximize the minimum individual SE of common message. The designed robust RS-based precoding can alleviate the negative impacts of asynchronous reception to a certain extent, especially when the asynchronous issues are moderate.
      It is highlighted that the uniform coverage characteristics of CF massive MIMO systems allow the RS with a simple low-complexity common precoding to achieve significant performance advantages.
\end{itemize}

Note that the conference version of this paper \cite{zheng2022performance} investigated the downlink performance of CF massive MIMO systems with asynchronous reception.
The rest of the paper is organized as follows. In Section \ref{se:model}, we describe the system model for capturing the joint effects caused by asynchronous reception and the basic principle of RS. Next, Section \ref{se:performance} presents the achievable downlink SE of the CF massive MIMO system with asynchronous reception for both coherent and non-coherent transmission schemes. Then, Section \ref{RS research} derives the sum SE expression for the RS-assisted CF massive MIMO system and the robust precoding for the common messages to reduce the effect of asynchronous reception. We provide numerical results and discussions in Section \ref{se:numerical}. Finally, Section \ref{se:conclusion} concludes this paper with a brief summary.

\textit{Notation:} We use boldface lowercase letters $\mathbf{x}$ and boldface uppercase letters $\mathbf{X}$ to represent column vectors and matrices, respectively.
The $n\times n$ identity matrix is ${{\mathbf{I}}_n}$.
Superscripts $x^\mathrm{*}$, $\mathbf{x}^\mathrm{T}$, and $\mathbf{x}^\mathrm{H}$ are used to denote conjugate, transpose, and conjugate transpose, respectively.
The absolute value, the Euclidean norm, the trace operator, and the definitions are denoted by $\left|  \cdot  \right|$, $\left\|  \cdot  \right\|$, ${\text{tr}}\left(  \cdot  \right)$, and $\triangleq$, respectively.
Finally, $x \sim \mathcal{C}\mathcal{N}\left( {0,{\sigma^2}} \right)$ represents a circularly symmetric complex Gaussian random variable $x$ with variance $\sigma^2$.

%\vspace{-2mm}

%----------------------------system model----------------------------
\section{System Model}\label{se:model}

\begin{figure}[t]
\vspace{0.2cm}
\centering
\includegraphics[scale=0.3]{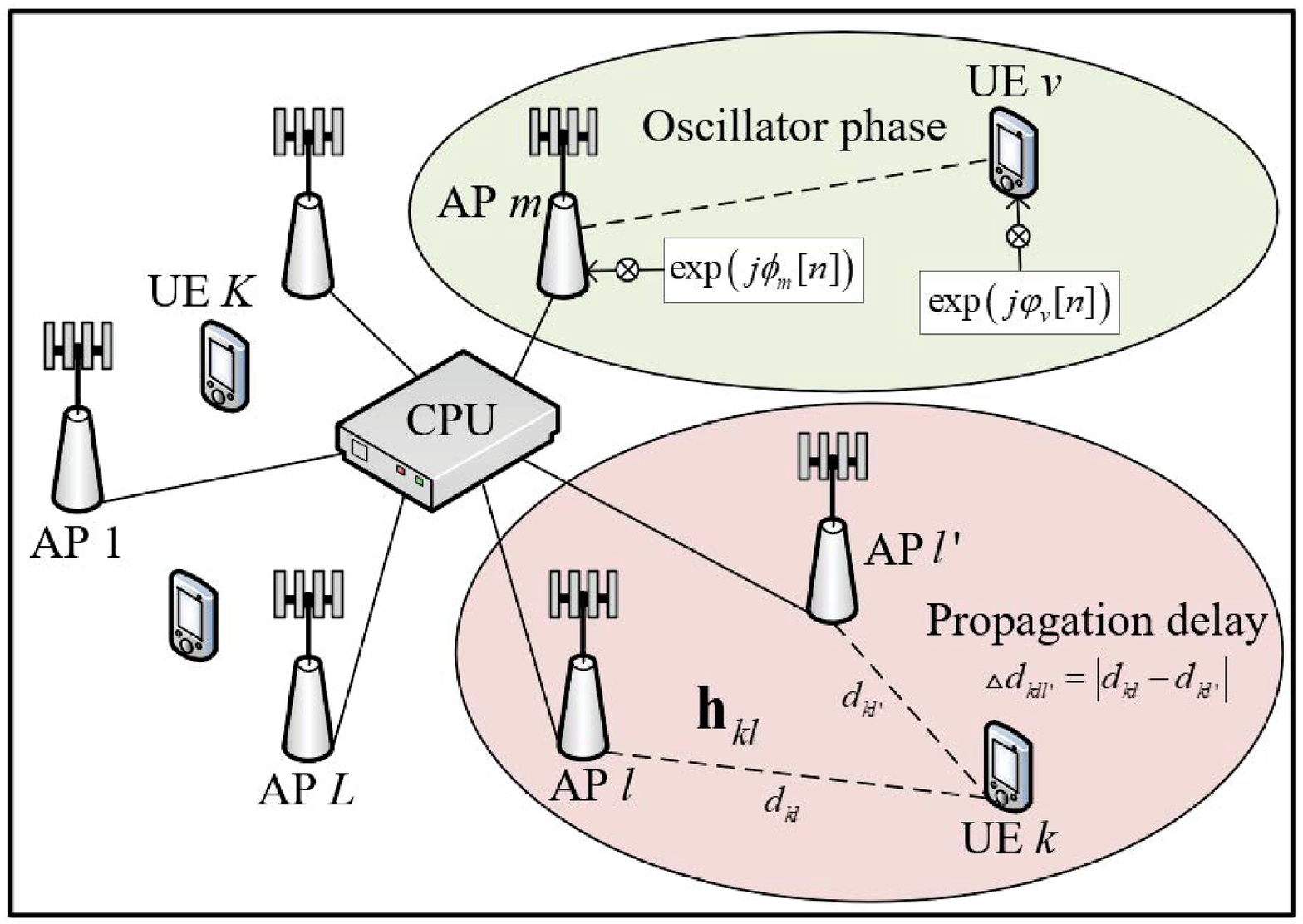}
\caption{Asynchronous reception in a CF massive MIMO system.} \vspace{-4mm}
\label{asyn}
\end{figure}

We consider a CF massive MIMO system comprising $L$ APs and $K$ UEs as illustrated in Fig.~\ref{asyn}. Besides, a single antenna and $N$ antennas are deployed for each UE and AP, respectively. It is assumed that all the $L$ APs simultaneously serve all the $K$ UEs via the same time-frequency resources. Moreover, we adopt the time-division duplex protocol with a standard coherence block model consisting of $\tau_c$ time instants (channel uses) with the uplink training phase occupying $\tau_p$ time instants and the downlink transmission phase assuming $\tau_c - \tau_p$ time instants. Besides, the frequency-flat channel between AP $l \in \left\{ {1, \ldots ,L} \right\}$ and UE $k \in \left\{ {1, \ldots ,K} \right\}$ at each coherence block is modeled as Rayleigh fading \cite{bjornson2019making}
\begin{align}\label{h_R}
{{\mathbf{h}}_{kl}} \sim \mathcal{C}\mathcal{N}\left( {{\mathbf{0}},{{\mathbf{R}}_{kl}}} \right) ,
\end{align}
where ${{\mathbf{R}}_{kl}} \in {\mathbb{C}^{N \times N}}$ represents the spatial correlation matrix and ${\beta _{kl}} \!\triangleq\! {\text{tr}}\left( {{{\mathbf{R}}_{kl}}} \right)/N$ denotes the large-scale fading coefficient.

\subsection{Asynchronous Reception}

The asynchronous reception of the wireless transceiver arises mainly from two factors: signal propagation delay differences and hardware oscillator errors. Specifically, these introduces a multiplicative random phase to the channel which are detailed as follows \footnote{Note that we mainly study the impact of asynchronous reception in terms of phase-asynchronization. However, performing coherent processing also requires synchronizations in time and frequency which has been widely studied in the literature \cite{etzlinger2018synchronization}.}.

\subsubsection{Delay Phase}

Due to the different physical positions of the APs in the CF architecture, the distances between each AP and a certain UE are different, resulting in asynchronous signal arrival to the UEs. This asynchronous reception introduces a constant phase shift as \cite{8676341}
\begin{align}
{\theta _{kl}} = {e^{ - j2\pi \frac{{{\Delta t_{kl}}}}{T_s}}} ,
\end{align}
where $\Delta t_{kl} = \Delta d_{kll'}/c$ is the timing offset of the signal intending for the $k$th UE and transmitted by the $l$th AP. Besides, $\Delta d_{kll'}$, $c$, and $T_s$ are the propagation delay, speed of light, and symbol duration, respectively. Without loss of generality, we assume that the first arrived signal to UE $k$ is from AP ${l'}$ and its timing offset is $\Delta t_{kl'} = 0$.
\subsubsection{Oscillator Phase}

Each AP and UE are assumed to be equipped with their own free-running oscillator \cite{9502552}, and the phase of the transmission symbol in each channel use is altered due to the phase noise. Then, the overall oscillator phase between AP $l$ and UE $k$ at each time instant can be defined by ${\vartheta _{kl}}\left[ n \right] \triangleq \exp \left( {j\left( {{\varphi _k}\left[ n \right] + {\phi _l}\left[ n \right]} \right)} \right)$ with the discrete-time Wiener phase model \cite{bjornson2015massive}
\begin{align}
{\varphi _k}\left[ n \right] &= {\varphi _k}\left[ {n - 1} \right] + \delta _k^{{\text{ue}}}\left[ n \right] , \\
{\phi _l}\left[ n \right] &= {\phi _l}\left[ {n - 1} \right] + \delta _l^{{\text{ap}}}\left[ n \right] ,
\end{align}
where ${\varphi _k}\left[ n \right]$ and ${\phi _l}\left[ n \right]$ are the oscillator phase of UE $k$ and AP $l$ at the $n$th time instant, respectively. Besides, the additive terms $\delta _k^{{\text{ue}}}\left[ n \right] \sim \mathcal{C}\mathcal{N}\left( {0,\sigma _{{k}}^2} \right)$ and $\delta _l^{{\text{ap}}}\left[ n \right] \sim \mathcal{C}\mathcal{N}\left( {0,\sigma _{{l}}^2} \right)$ are the phase increment of AP $l$ and UE $k$ at the $n$th time instant. Note that $\sigma _i^2 = 4{\pi ^2}{f_c^2}{c_i}{T_s},i = {k}, {l} $, denote the phase increment variance \cite{bjornson2015massive}, where $f_c$ is the carrier frequency and $c_i$ is a constant dependent on the oscillator.
\begin{rem}
Note that we focus on the scenario which each AP and UE has its oscillator such that their oscillator phase processes are considered as mutually independent. For the case of analysis, we assume independent and identically distributed oscillator phase statistics across different APs and UEs, i.e., $\sigma _{{k}}^2 = \sigma _{\mathrm{ue}}^2$ and $\sigma _{{l}}^2 = \sigma _{\mathrm{ap}}^2$, $\forall k,l$.
\end{rem}
Considering the effect of both the delay and oscillator phases, the channel between the $k$th UE and the $l$th AP at the $n$th time instant is expressed as
\begin{align}\label{gkl}
{{\mathbf{g}}_{kl}}\left[ n \right] = {\theta _{kl}}{{\mathbf{h}}_{kl}}\left[ n \right] = {\theta _{kl}}{\vartheta _{kl}}\left[ n \right]{{\mathbf{h}}_{kl}} ,n = 1, \ldots ,{\tau _c} ,
\end{align}
where ${{\mathbf{h}}_{kl}}\left[ n \right] \in {\mathbb{C}^{N \times 1}}$ is the channel that combines the oscillator phase and it is random in each time instant. Besides, the parameter ${\theta _{kl}}$ is mainly determined by the positions of APs and UEs, and thus can be considered as a constant among multiple coherent blocks\footnote{We assume that the delay phase can be perfectly known by positioning or other technologies. However, for the design of downlink precoding in the sequel, we will consider both the cases of with or without (used/forgotten) delay phase to quantify its impact \cite{9531352}.}.

%\newgeometry{left = 0.6in, bottom=0.96in}
\subsection{Channel Estimation}

\begin{figure}[t]
\centering
\includegraphics[scale=0.55]{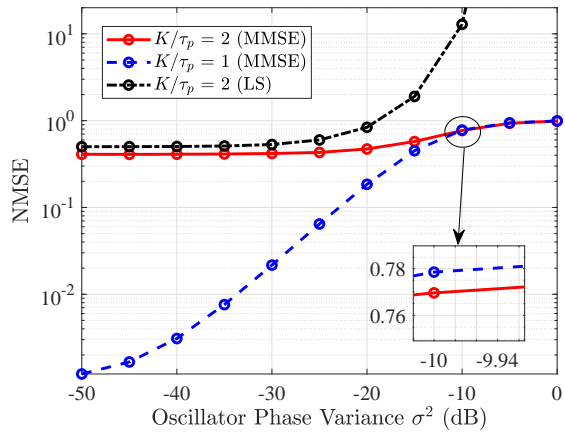}
\caption{The NMSE of MMSE and LS channel estimation with different oscillator phase variances at $\text{SNR} = 30$ dB ($L=100$, $N=2$, $\sigma_\text{ap}^2=\sigma_\text{ue}^2=\sigma^2$).} \vspace{-4mm}
\label{figure1}
\end{figure}

We assume $\tau_p$ mutually orthogonal time-multiplexed pilot sequences are employed \cite{9322468}, where the different sequences are temporally orthogonal since the pilot sequence $t$ equates to transmitting a pilot signal only at time instant $t$ \cite{bjornson2015massive}.
This can also be seen that the pilot sequence $t$ is a $\tau_p$-dimensional vector with all zeros except one in the $t$th channel use.
Besides, a large-scale network with $K>\tau_p$ is studied such that different UEs may use the same time instant for channel estimation. Moreover, the index of the time instant allocated to UE $k$ is denoted by ${t_k} \in \left\{ {1, \ldots ,{\tau _p}} \right\}$, and the other UEs that exploit the same time instant for pilot transmission as UE $k$ is defined by ${\mathcal{P}_k} = \left\{ {i:{t_i} = {t_k}} \right\} \in \{ 1, \ldots ,K\} $.
Considering the effect of asynchronous reception, the received signal at AP $l$ from UE $k$ at time instant $t_k$ is given by
\begin{align}\label{zl1}
{{\mathbf{z}}_l}\left[ {{t_k}} \right] = \sum\limits_{i \in {\mathcal{P}_k}} {\sqrt {{p_i}} {{\mathbf{g}}_{il}}\left[ {{t_i}} \right] + {{\mathbf{w}}_l}\left[ {{t_k}} \right]},
\end{align}
where $p_i $ denotes the pilot transmit power of UE $i$ and ${{\mathbf{w}}_l}\left[ {{t_k}} \right] \sim \mathcal{C}\mathcal{N}\left( {0,{\sigma ^2}{{\mathbf{I}}_N}} \right)$ represents the receiver noise at AP $l$ with the noise power $\sigma ^2$.
This received signal is exploited to estimate (or predict) the channel realization at any time instant in the block. Yet, the accuracy of the channel estimate is degraded as the temporal gap between the considered channel realization and the pilot transmission grows.
Without lost of generality, we consider the estimates of the channels at time instant $\tau_p + 1$.
In addition, $\lambda = \tau_p + 1$ is defined to simplify the notation and \eqref{zl1} can be rewritten as
\begin{align}\label{zltk}
{{\mathbf{z}}_l}\left[ {{t_k}} \right] = \sqrt {{p_k}} {\theta _{kl}}\Theta _{kl}^*\left[ \lambda - t_k  \right]{{\mathbf{h}}_{kl}}\left[ \lambda  \right] + \sum\limits_{i \in {\mathcal{P}_k}/\left\{ k \right\}} {\sqrt {{p_i}} {\theta _{il}}{{\mathbf{h}}_{il}}\left[ {{t_i}} \right] + {{\mathbf{w}}_l}\left[ {{t_k}} \right]} ,
\end{align}
where ${\Theta _{kl}}\left[ \lambda \!-\! t_k \right]$ denotes the phase increment from time instant $t_k$ to time instant $\lambda$, defined as
\begin{align}\label{theta1}
{\Theta _{kl}}\left[ \lambda - t_k \right] = {\vartheta _{kl}}\left[ \lambda \right]\vartheta _{kl}^*\left[ t_k  \right] = \exp \left( {j\sum\limits_{s = t_k  + 1}^\lambda {\left( {\delta _k^{{\text{ue}}}\left[ s \right] + \delta _l^{{\text{ap}}}\left[ s \right]} \right)} } \right) .
\end{align}
By exploiting the characteristic function of Gaussian random variable, the mean of \eqref{theta1} is
\begin{align}\label{theta2}
\mathbb{E}\left\{ {{\Theta _{kl}}\left[ \lambda - t_k \right]} \right\} = {e^{ - \frac{{\lambda  - {t_k}}}{2}\left( {\sigma _{{\text{ap}}}^2 + \sigma _{{\text{ue}}}^2} \right)}} .
\end{align}
Then, applying the standard minimum mean square error (MMSE) estimation \cite{Ngo2017Cell}, the MMSE estimate ${{{\mathbf{\hat h}}}_{kl}}\left[ \lambda  \right]$ of the channel coefficient ${{\mathbf{h}}_{kl}}\left[ \lambda  \right]$ can be computed by each AP $l$ as
\begin{align}\label{hhat}
  {{{\mathbf{\hat h}}}_{kl}}\left[ \lambda  \right] &  =  \frac{{\mathbb{E}\left\{ {{{\mathbf{h}}_{kl}}\left[ \lambda  \right]{\mathbf{z}}_l^{\text{H}}\left[ {{t_k}} \right]} \right\}}}{{\left( {\mathbb{E}\left\{ {{{\mathbf{z}}_l}\left[ {{t_k}} \right]{\mathbf{z}}_l^{\text{H}}\left[ {{t_k}} \right]} \right\}} \right)}}{{\mathbf{z}}_l}\left[ {{t_k}} \right] \mathop  = \limits^{\left( a \right)} \frac{{\sqrt {{p_k}} \theta _{kl}^*\mathbb{E}\left\{ {{\Theta _{kl}}\left[ {\lambda  - {t_k}} \right]} \right\}\mathbb{E}\left\{ {{{\mathbf{h}}_{kl}}\left[ \lambda  \right]{\mathbf{h}}_{kl}^{\text{H}}\left[ \lambda  \right]} \right\}}}{{\mathbb{E}\left\{ {{{\mathbf{z}}_l}\left[ {{t_k}} \right]{\mathbf{z}}_l^{\text{H}}\left[ {{t_k}} \right]} \right\}}}{{\mathbf{z}}_l}\left[ {{t_k}} \right] \notag \\
   &\mathop  = \limits^{\left( b \right)} \frac{{\sqrt {{p_k}} {e^{ - \frac{{\lambda  - {t_k}}}{2}\left( {\sigma _{{\text{ap}}}^2 + \sigma _{{\text{ue}}}^2} \right)}}\theta _{kl}^{\text{*}}{{\mathbf{R}}_{kl}}}}{{\sum\limits_{i \in {\mathcal{P}_k}} {{p_i}{{\mathbf{R}}_{il}}}  + {\sigma ^2}{{\mathbf{I}}_N}}}{{\mathbf{z}}_l}\left[ {{t_k}} \right],
\end{align}
where $\left(a\right)$ follows from the independence between phases and channels, and from that channels between different UEs are independent of each other. $\left(b\right)$ follows from computing the variances of the channel \eqref{h_R} and the received signal \eqref{zltk}, and from using the fact in \eqref{theta2}.
In addition, the distribution of the estimate ${{{\mathbf{\hat h}}}_{kl}}\left[ \lambda  \right]$ and the estimation error ${{{\mathbf{\tilde h}}}_{kl}}\left[ \lambda  \right] = {{\mathbf{h}}_{kl}}\left[ \lambda  \right] - {{{\mathbf{\hat h}}}_{kl}}\left[ \lambda  \right]$ are $\mathcal{C}\mathcal{N}\left( {{\mathbf{0}},{{\mathbf{Q}}_{kl}}} \right)$ and $ \mathcal{C}\mathcal{N}\left( {{\mathbf{0}},{{\mathbf{R}}_{kl}}-{{\mathbf{Q}}_{kl}}} \right)$, where
\begin{align}\label{Qkl}
{{\mathbf{Q}}_{kl}} = {p_k}{e^{ - \left( {\lambda  - {t_k}} \right)\left( {\sigma _{{\text{ap}}}^2 + \sigma _{{\text{ue}}}^2} \right)}}{{\mathbf{R}}_{kl}}{{\mathbf{\Psi }}_{kl}}{{\mathbf{R}}_{kl}} ,
\end{align}
with
\begin{align}
{{\mathbf{\Psi }}_{kl}} = {\left( {\sum\limits_{i \in {\mathcal{P}_k}} {{p_i}{{\mathbf{R}}_{il}}}  + {\sigma ^2}{{\mathbf{I}}_N}} \right)^{ - 1}} .
\end{align}
Moreover, to simplify the notation, we define
\begin{align}
{{{\mathbf{\bar Q}}}_{kil}} = \sqrt {{p_k}{p_i}} {e^{ - \left( {\lambda  - {t_k}} \right)\left( {\sigma _{{\text{ap}}}^2 + \sigma _{{\text{ue}}}^2} \right)}}{{\mathbf{R}}_{il}}{{\mathbf{\Psi }}_{kl}}{{\mathbf{R}}_{kl}} ,
\end{align}
that represents the covariance matrix of the estimate for UE $k$ and UE $i$ at AP $l$.
Note that each UE is assumed to be aware of the channel statistics and the signal detection is performed with the required channel distribution information available.
\begin{rem}
To compare the estimation quality under different degrees of asynchronous reception, we utilize the normalized mean square error (NMSE) given as
\begin{align}\label{NMSE}
{\mathrm{NMSE}}{_{kl}} = \frac{{{\mathrm{tr}}\left( {{{\mathbf{R}}_{kl}} - {{\mathbf{Q}}_{kl}}} \right)}}{{{\mathrm{tr}}\left( {{{\mathbf{R}}_{kl}}} \right)}} , \forall k,l,
\end{align}
which is an appropriate metric to measure the relative estimation error. Note that ${{{\mathbf{\hat h}}}_{kl}}\left[ \lambda  \right]$ and ${{{\mathbf{\tilde h}}}_{kl}}\left[ \lambda  \right]$ are independent due to the properties of MMSE estimation. Therefore, the values of NMSE are between 0 and 1, which denote perfect and extremely impaired estimation, respectively.
Following the similar steps as in \cite[Section III]{bjornson2017massive}, we derive the NMSE of the least-squares (LS) estimator for comparison as
\begin{align}
{\mathrm{NMS}}{{\mathrm{E}}^\mathrm{ls}_{kl}} = \frac{{{e^{\left( {\lambda  - {t_k}} \right)\left( {\sigma _{{\mathrm{ap}}}^2 + \sigma _{{\mathrm{ue}}}^2} \right)}}{\mathrm{tr}}\left( {{\mathbf{\Psi }}_{kl}^{ - 1}} \right)}}{{{p_k}{\mathrm{tr}}\left( {{{\mathbf{R}}_{kl}}} \right)}} - 1 , \forall k,l .
\end{align}
Note that since the estimate and the estimation error of LS are correlated, the value of NMSE can be higher than one.
\end{rem}
As illustrated in Fig.~\ref{figure1}, we compare the NMSE of MMSE and LS channel estimations with different degrees of oscillator phase as the $\text{SNR} = 30$ dB for all the links. It is clear that the MMSE estimation quality gets worse with the increasing of the oscillator phase variance. The reason is that the random oscillator phase destroys the orthogonality of the received pilot signal. We also find that the NMSE is sensitive to the oscillator phase when the number of pilots are sufficient. In some extreme asynchronous cases, exploiting non-orthogonal pilots can achieve a more accurate estimation. This result can be explained by \eqref{Qkl} and \eqref{NMSE} as that pilot contamination makes the value of NMSE worse and change slowly.
It is also found that the estimation accuracy of LS is generally lower than that of MMSE and the LS estimator performs very poorly when the oscillator phase variances are large.

\subsection{Rate-Splitting Strategy}

\begin{figure}[t]
\centering
\includegraphics[scale=0.45]{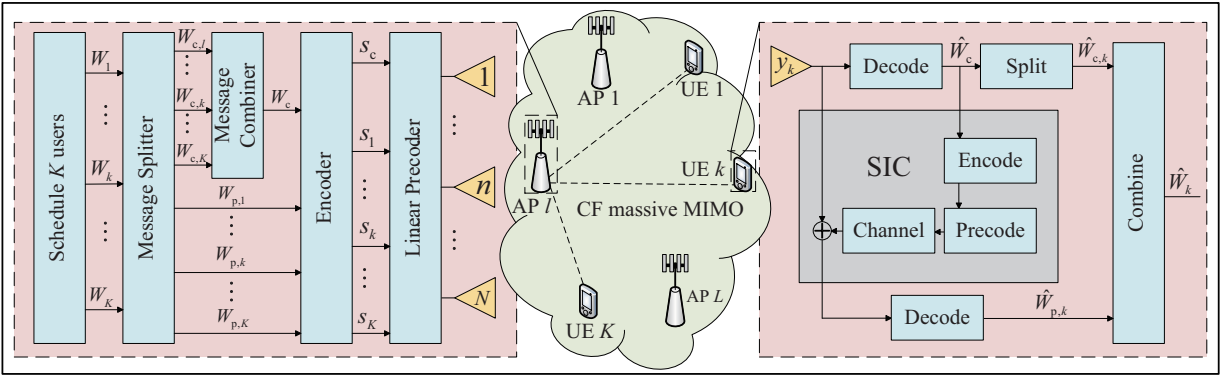}
\caption{A RS-assisted CF massive MIMO system \cite{7470942}.\ \ \ \ \ \ \ \ \ \ \ \ \ \ \ \ \ \ \ \ \ \ \ \ \ \ \ \ \ \ \ \ \ \ \ \ \ \ \ \ \ \ \ \ \ \ \ \ \ \ \ \ \ \ \ } \vspace{-4mm}
\label{RS}
\end{figure}

Since imperfect CSI is inevitable in the considered CF system due to asynchronous issues, we propose the use of RS for the downlink to compensate the possible performance degradation.
The rationale behind the RS strategy for multi-user downlink is to first split the message for each UE into a private message and a common message, then combine all the common sub-messages into a super common message, and finally all the messages are simultaneously transmitted by means of superposition coding \cite{7470942}.
The principle of RS is shown in Fig.~\ref{RS}, specifically, the message $W_k \in {\mathbb{C}}$ intended for UE $k$ at the transmitter is split into a private part $W_{\text{p},k} \in {\mathbb{C}}$ and a common part $W_{\text{c},k} \in {\mathbb{C}}$. The private parts ${W_{{\text{p,1}}}}, \ldots ,{W_{{\text{p}},K}}$ are independently encoded into the private streams ${s_1}, \ldots ,{s_K}$, respectively, and the common parts of all the users, ${W_{{\text{c,1}}}}, \ldots ,{W_{{\text{c}},K}}$, are combined into a common message $W_\text{c}$, which is encoded into a common stream $s_\text{c}$ using a public codebook. Then, these $K+1$ streams are transmitted after linear precoding.
At the receivers, each UE firstly decodes the common stream by treating all the private streams as noise. Then, SIC is adopted to remove the decoded common stream from the received signal under the assumption of error-free decoding\footnote{It is tolerable to assume error-free decoding to obtain preliminary analytical results because the considered one-layer RS scheme requires each UE executes SIC only once \cite{mao2018rate}.} \cite{7434643}. Each user then decodes its private stream by treating other private streams as noise.
%It is necessary to investigate the effect of imperfect SIC on RS-assisted CF massive MIMO systems, however this is left for future works by modeling the error propagation factor as a chi-squared random variable \cite{9115309}.

Utilizing the common precoding vector ${{\mathbf{v}}_{\text{c},l}} \in {\mathbb{C}^{N \times 1}}$ and the private precoding vector ${{\mathbf{v}}_{kl}} \in {\mathbb{C}^{N \times 1}}$ in asynchronous CF massive MIMO systems with RS, the transmitted signal from AP $l$ at the $n$th time instant is given by
\begin{align}\label{data_RS}
{{\mathbf{x}}_l}\left[ n \right] = \sqrt {{p_{{\text{dc}}}}{\eta _l}} {{\mathbf{v}}_{{\text{c}},l}}{s_{\text{c}}}\left[ n \right] + \sqrt {{p_{{\text{dp}}}}{\mu _l}} \sum\limits_{i = 1}^K {{{\mathbf{v}}_{il}}{s_i}\left[ n \right]},
\end{align}
where $s_\text{c} \in {\mathbb{C}}$ is the common message and $s_i \in {\mathbb{C}}$ is the private message of UE $i$. Besides, $p_{\text{dc}}$ and $p_{\text{dp}}$ are the powers allocated to the common and private message, respectively. Moreover, $\eta_l$ and $\mu_l$ are the power normalization parameters for the common and private precoding, respectively, and are respectively denoted by
\begin{align}
\label{etal} {\eta _l} & \triangleq \frac{1}{{\mathbb{E}\left\{ {{{\left\| {{{\mathbf{v}}_{{\text{c}},l}}} \right\|}^2}} \right\}}} , \\
\label{mul} {\mu _l} &\triangleq \frac{1}{{\sum\limits_{i = 1}^K {\mathbb{E}\left\{ {{{\left\| {{{\mathbf{v}}_{il}}} \right\|}^2}} \right\}} }}.
\end{align}
We assume that the downlink maximum transmission power of each AP is the same represented by $p_\text{d}$ \cite{Ngo2017Cell}. Besides, the power splitting factor $\rho \left(0 \leqslant \rho \leqslant 1\right)$ is defined to adjust the fraction of the power allocated to the transmission of the common messages at each AP\footnote{Due to the small channel differences in CF massive MIMO networks, we assume that the power splitting factors are identical at all APs to arrive at a preliminary conclusion. In the future work, we will design the optimal power splitting factor for each AP to further improve the system performance.}. Hence, we have $p_\text{dc} = \rho p_\text{d}$ and $p_\text{dp} = \left(1 - \rho\right) p_\text{d}$. In Section~\ref{RS research}, we will introduce the binary search method to determine the optimal power splitting factor. When the power splitting factor $\rho = 0$, RS-assisted CF massive MIMO systems degenerate into the conventional CF massive MIMO.

\begin{rem}
The RS strategy is promising in multi-user transmission with imperfect CSI \cite{7892949}. In contrast to the conventional strategy that treats any multi-user interference originating from the imperfect CSI as noise, the RS strategy has the ability to treat the interference as noise and perform interference decoding through the presence of a common message. This ability of the decoding part of the interference is the key for boosting the sum SE performance.
\end{rem}

\section{Downlink Asynchronous Transmission}\label{se:performance}

In this part, we first focus on the effect of asynchronous reception on CF massive MIMO systems. Therefore, we assume the power splitting factor $\rho = 0$ for a conventional approach\footnote{Note that the obtained results for $\rho=0$ will serve as a building block for the study of RS-assisted CF massive MIMO systems in Section~\ref{RS research}.}, which makes $p_\text{dc} = 0$ and $p_\text{dp} = p_\text{d}$. Then, considering both the coherent and non-coherent downlink transmission, we adopt both the delay phase used (DU) and delay phase forgotten (DF) maximum-ratio (MR) private precoding to derive the closed-form SE performance expressions.

\subsection{Coherent Downlink Transmission}

We assume that the CF massive MIMO applies the coherent joint transmission for the downlink transmission, which means that each AP sends the same data symbol to each UE as the other APs \cite{8809413}.
When the power splitting factor $\rho = 0$ (without RS), the transmitted signal in \eqref{data_RS} reduces to
\begin{align}\label{x_l}
{{\mathbf{x}}_l}\left[ n \right] = \sqrt {{p_{{\text{dp}}}}{\mu _l}} \sum\limits_{i = 1}^K {{{\mathbf{v}}_{il}}{s_i}\left[ n \right]}.
\end{align}
Then, the received signal of the $k$th UE at the $n$th time instant is
\begin{align}\label{resignal}
y_k^{{\text{p,co}}}\left[ n \right] = \sqrt {{p_{\text{dp}}}} \sum\limits_{l = 1}^L {{\mathbf{g}}_{kl}^{\text{H}}\left[ n \right]\sqrt {{\mu _l}} {{\mathbf{v}}_{kl}}{s_k}\left[ n \right]} + \sqrt {{p_{\text{dp}}}} \sum\limits_{i \ne k}^K {\sum\limits_{l = 1}^L {{\mathbf{g}}_{kl}^{\text{H}}\left[ n \right]\sqrt {{\mu _l}} {{\mathbf{v}}_{il}}{s_i}\left[ n \right]} }  + {w_k}\left[ n \right] ,
\end{align}
where ${w_k}\left[ n \right] \sim \mathcal{C}\mathcal{N}\left( {0,\sigma _{\text{d}}^2} \right)$ denotes the receiver noise at UE $k$.
For minimizing the MSE, ${\text{MS}}{{\text{E}}_{kl}} = \mathbb{E}\left\{ {{{\left| {{s_k}\left[ n \right] - y_k^{{\text{p,co}}}\left[ n \right]\left( {{{\mathbf{v}}_{kl}}} \right)} \right|}^2}\left| {{{{\mathbf{\hat h}}}_{il}}\left[ \lambda  \right]} \right.} \right\}$, we derive the local DU-MMSE private precoding vector as
\begin{align}
{{\mathbf{v}}_{kl}} = {\theta _{kl}}{p_{{\text{dp}}}}{\left( {\sum\limits_{i = 1}^K {{p_{{\text{dp}}}}\left( {{{{\mathbf{\hat h}}}_{il}}\left[ \lambda  \right]{\mathbf{\hat h}}_{il}^{\text{H}}\left[ \lambda  \right] + {{\mathbf{C}}_{il}}} \right) + {\sigma ^2}{{\mathbf{I}}_N}} } \right)^{ - 1}}{{{\mathbf{\hat h}}}_{kl}}\left[ \lambda  \right],
\end{align}
where ${{\mathbf{C}}_{il}} = {{\mathbf{R}}_{il}} - {{\mathbf{Q}}_{il}}$ is the variance of the estimation error. Moreover, low-complexity MR private precoding is exploited to obtain analytical results, which does not affect the validity of our conclusions.
In the following, we introduce a theorem to study the system performance.

\begin{thm}\label{thm1}
With the help of the channel estimate \eqref{hhat} and the received signal \eqref{resignal}, using the DU-MR private precoding ${{\mathbf{v}}_{kl}} = {\theta _{kl}}{{{\mathbf{\hat h}}}_{kl}}\left[ \lambda  \right]$, the downlink achievable rate of UE $k$ is lower bounded as
\begin{align}
{\mathrm{SE}}_k^{{\mathrm{p,co,du}}} = \frac{1}{{{\tau _c}}}\sum\limits_{n = \lambda }^{{\tau _c}} {{{\log }_2}\left( {1 + {\mathrm{SINR}}_k^{{\mathrm{p,co,du}}}\left[ n \right]} \right)} ,
\end{align}
with
\begin{align}\label{SINR_CF}
{\mathrm{SINR}}_k^{{\mathrm{p,co}},{\mathrm{du}}}\left[ n \right] = \frac{{ \eta _n^{{\mathrm{ap}}}\eta _n^{{\mathrm{ue}}} {p_{\mathrm{dp}}}{{\left| {\sum\limits_{l = 1}^L {\sqrt {{\mu _l}} {\mathrm{tr}}\left( {{{\mathbf{Q}}_{kl}}} \right)} } \right|}^2}}}{{{p_{\mathrm{dp}}}\Xi _k^{{\mathrm{p,co}},{\mathrm{du}}}\left[ n \right] - \eta _n^{{\mathrm{ap}}}\eta _n^{{\mathrm{ue}}}{p_{\mathrm{dp}}}{\left| {\sum\limits_{l = 1}^L {\sqrt {{\mu _l}} {\mathrm{tr}}\left( {{{\mathbf{Q}}_{kl}}} \right)} } \right|^2} + \sigma _{\mathrm{d}}^2}} ,
\end{align}
where
\begin{align}
   {\Xi _k^{{\mathrm{p,co}},{\mathrm{du}}}\left[ n \right]} &= \sum\limits_{i = 1}^K {\sum\limits_{l = 1}^L {{\mu _l}{\mathrm{tr}}\left( {{{\mathbf{Q}}_{il}}{{\mathbf{R}}_{kl}}} \right)}} \notag\\
   &  + \left( {1 - \eta _n^{{\mathrm{ap}}} } \right)\sum\limits_{i \in {\mathcal{P}_k}}^K {\sum\limits_{l = 1}^L {{\mu _l}{{\left| {{\mathrm{tr}}\left( {{{{\mathbf{\bar Q}}}_{kil}}} \right)} \right|}^2}} }  +\eta _n^{{\mathrm{ap}}}\sum\limits_{i \in {\mathcal{P}_k}}^K {{{\left| {\sum\limits_{l = 1}^L {\sqrt {{\mu _l}} {\mathrm{tr}}\left( {{{{\mathbf{\bar Q}}}_{kil}}} \right)} } \right|}^2}} .
\end{align}
Note that we have $\eta _n^{{\mathrm{ap}}} \triangleq {e^{ - \left( {n - \lambda } \right)\sigma _{{\mathrm{ap}}}^2}}$ and $\eta _n^{{\mathrm{ue}}} \triangleq {e^{ - \left( {n - \lambda } \right)\sigma _{{\mathrm{ue}}}^2}}$.
\end{thm}
\begin{IEEEproof}
See Appendix A.
\end{IEEEproof}

\begin{cor}\label{appro}
Only keeping the oscillator phase parameters, the approximate SINR expression of \eqref{SINR_CF} under which the number of antennas tends to infinity $\left(LN \to \infty \right)$ can be derived as $1/\left( {1/\left(L\eta _n^{{\mathrm{ap}}}\eta _n^{{\mathrm{ue}}}\right) + \left( {L - 1} \right)/\left(L\eta _n^{{\mathrm{ue}}}\right) + a} \right)$, where $a$ is a constant. It is clear that the SINR decreases as $\sigma_\mathrm{ap}^2$ and $\sigma_\mathrm{ue}^2$ increase, and $\sigma_\mathrm{ue}^2$ has a larger effect on the SINR compared with $\sigma_\mathrm{ap}^2$. Moreover, increasing the number of APs can only significantly reduce the influence of the oscillator phase at the APs.
\end{cor}
From \eqref{SINR_CF}, we can find that the delay phase has no impact on the SINR expression when the DU-MR private precoding is adopted. To study and characterize the influence of the delay phase on the system, we investigate the SE performance using DF-MR private precoding with ${{\mathbf{v}}_{kl}} = {{{\mathbf{\hat h}}}_{kl}}\left[ \lambda  \right]$. Following the similar steps as in Theorem~\ref{thm1}, we obtain
\begin{align}\label{SINR_CF_DU}
{\text{SINR}}_k^{{\text{p,co}},{\text{df}}}\left[ n \right] = \frac{{ \eta _n^{{\mathrm{ap}}}\eta _n^{{\mathrm{ue}}} {p_{\text{dp}}}{{\left| {\sum\limits_{l = 1}^L {{\theta _{kl}^*}\sqrt {{\mu _l}} {\text{tr}}\left( {{{\mathbf{Q}}_{kl}}} \right)} } \right|}^2}}}{{{p_{\text{dp}}}\Xi _k^{{\text{p,co}},{\text{df}}}\left[ n \right] - \eta _n^{{\mathrm{ap}}}\eta _n^{{\mathrm{ue}}}{p_{\text{dp}}}{\left| {\sum\limits_{l = 1}^L {{\theta _{kl}^*}\sqrt {{\mu _l}} {\text{tr}}\left( {{{\mathbf{Q}}_{kl}}} \right)} } \right|^2}  + \sigma _{\text{d}}^2}} ,
\end{align}
where
\begin{align}
   {\Xi _k^{{\text{p,co}},{\text{df}}}\left[ n \right]} &= \sum\limits_{i = 1}^K {\sum\limits_{l = 1}^L {{\mu _l}{\text{tr}}\left( {{{\mathbf{Q}}_{il}}{{\mathbf{R}}_{kl}}} \right)} }    \notag \\
   & + \left( {1 - \eta _n^{{\mathrm{ap}}}} \right)\sum\limits_{i \in {\mathcal{P}_k}}^K {\sum\limits_{l = 1}^L {{\mu _l}{{\left| {{\text{tr}}\left( {{{{\mathbf{\bar Q}}}_{kil}}} \right)} \right|}^2}} } + \eta _n^{{\mathrm{ap}}}\sum\limits_{i \in {\mathcal{P}_k}}^K {{{\left| {\sum\limits_{l = 1}^L {{\theta _{il}^*}\sqrt {{\mu _l}} {\text{tr}}\left( {{{{\mathbf{\bar Q}}}_{kil}}} \right)} } \right|}^2}} .
\end{align}
Comparing \eqref{SINR_CF} and \eqref{SINR_CF_DU}, it is clear that the propagation delay caused by different physical positions of the APs introduces random phases to the desired signal, which will reduce the SE performance of systems.

\subsection{Non-Coherent Downlink Transmission}

For alleviating the phase-synchronization requirements of the APs imposed by coherent transmission, we apply a non-coherent joint transmission in the downlink CF massive MIMO such that each AP sends the different data symbol to each UE than the other APs \cite{8809413}.
Therefore, the transmitted signal from AP $l$ at time instant $n$ is expressed as
\begin{align}
{{\mathbf{x}}_l}\left[ n \right] = \sqrt {{p_{\text{dp}}}{\mu _l}} \sum\limits_{i = 1}^K {{{\mathbf{v}}_{il}}{s_{il}}\left[ n \right]} ,
\end{align}
where ${s_{il}}\left[ n \right] \in {\mathbb{C}}$ is the symbol sent to UE $i$ which is different for all the APs.
Then, the received signal of the $k$th UE at the $n$th time instant is
\begin{align}\label{resignal1}
y_k^{{\text{p,nc}}}\left[ n \right] = \sqrt {{p_{\text{dp}}}} \sum\limits_{l = 1}^L {{\mathbf{g}}_{kl}^{\text{H}}\left[ n \right]\sqrt {{\mu _l}} {{\mathbf{v}}_{kl}}{s_{kl}}\left[ n \right]} + \sqrt {{p_{\text{dp}}}} \sum\limits_{i \ne k}^K {\sum\limits_{l = 1}^L {{\mathbf{g}}_{kl}^{\text{H}}\left[ n \right]\sqrt {{\mu _l}} {{\mathbf{v}}_{il}}{s_{il}}\left[ n \right]} }  + {w_k}\left[ n \right] .
\end{align}
Note that the $k$th UE needs to employ SIC after receiving the signals from all the $L$ APs in order to detect the signals sent by the different APs \cite{8809413}. Specifically, the UE first detects the signal received from the first AP and the remaining signal is treated as interference. By that analogy, the UE detects the signal transmitted by the $l$th AP and considers the signal sent from the $(l + 1)$th AP to the $L$th AP as interference, thereby detecting the signal ${s_{kl}}\left[ n \right]$. Keep in mind that the SE is not affected by the APs' relative order of decoding.
However, a specific decoding order must be chosen before the individual signals may be encoded \cite{9322468,8809413}.

\begin{thm}\label{thm2}
Based on the received signal \eqref{resignal1} and using the DU-MR private precoding ${{\mathbf{v}}_{kl}} = {\theta _{kl}}{{{\mathbf{\hat h}}}_{kl}}\left[ \lambda  \right]$, the downlink achievable rate of UE $k$ is lower bounded as
\begin{align}
{\mathrm{SE}}_k^{{\mathrm{p,nc,du}}} = \frac{1}{{{\tau _c}}}\sum\limits_{n = \lambda }^{{\tau _c}} {{{\log }_2}\left( {1 + {\mathrm{SINR}}_k^{{\mathrm{p,nc,du}}}\left[ n \right]} \right)} ,
\end{align}
with
\begin{align}\label{SINR_CF_NC}
{\mathrm{SINR}}_k^{{\mathrm{p,nc,du}}}\left[ n \right] = \frac{{ \eta _n^{{\mathrm{ap}}}\eta _n^{{\mathrm{ue}}} {p_{\mathrm{dp}}} \sum\limits_{l = 1}^L {{\mu _l}} {{\left| {{\mathrm{tr}}\left( {{{\mathbf{Q}}_{kl}}} \right)} \right|}^2}}}{{ {p_{\mathrm{dp}}}\Xi _k^{{\mathrm{p,nc}},{\mathrm{du}}}\left[ n \right]  - \eta _n^{{\mathrm{ap}}}\eta _n^{{\mathrm{ue}}} {p_{\mathrm{dp}}} \sum\limits_{l = 1}^L {{\mu _l}} {{\left| {{\mathrm{tr}}\left( {{{\mathbf{Q}}_{kl}}} \right)} \right|}^2} + {\sigma ^2_\mathrm{d}}}} ,
\end{align}
where
\begin{align}
\Xi _k^{{\mathrm{p,nc}},{\mathrm{du}}}\left[ n \right] = \sum\limits_{i = 1}^K {\sum\limits_{l = 1}^L {{\mu _l}} {\mathrm{tr}}\left( {{{\mathbf{Q}}_{il}}{{\mathbf{R}}_{kl}}} \right)}  + \sum\limits_{i \in {\mathcal{P}_k}}^K {\sum\limits_{l = 1}^L {{\mu _l}} {{\left| {{\mathrm{tr}}\left( {{{{\mathbf{\bar Q}}}_{kil}}} \right)} \right|}^2}} .
\end{align}
\end{thm}
\begin{IEEEproof}
Follow the SIC operation for non-coherent transmission in \cite[Appendix C]{9322468} and the similar steps for proving Theorem \ref{thm1}.
\end{IEEEproof}

\begin{rem}\label{sinr_nc}
Note that if the DF-MR private precoding is used, we can also derive ${\mathrm{SIN}}{{\mathrm{R}}^{\mathrm{p,nc,df}}_k}\left[ n \right]$ having the same SINR expression as \eqref{SINR_CF_NC}. Therefore, we conclude that the non-coherent transmission can effectively overcome the influence of asynchronous reception, and even eliminate the influence of delay phase, at the expense of poor SE.
In future CF networks, a trade-off between high performance and strong robustness may be achieved by a hybrid scheme that includes both coherent and non-coherent transmission schemes.
\end{rem}

\subsection{Power Control for Private Messages}

Moreover, various power control methods can be adopted in our systems to further improve the system performance. To this end, we rewrite \eqref{x_l} as
\begin{align}
{{\mathbf{x}}_l}\left[ n \right] = \sqrt {{p_{{\text{dp}}}}} \sum\limits_{i = 1}^K {\sqrt {{\mu _{il}}} {{\mathbf{v}}_{il}}{s_i}\left[ n \right]},
\end{align}
where $\mu _{il}\geqslant 0$ is the power control coefficients chosen to satisfy the downlink power constraint as $\mathbb{E}\left\{ {{{\left| {{{\mathbf{x}}_l}\left[ n \right]} \right|}^2}} \right\} \leqslant {p_{{\text{dp}}}}$. With the help of the statistical channel cooperation power control in our previous work \cite{9322468}, we derive
\begin{align}
{\mu _{kl}} = \frac{{\bar \beta _k^{\alpha} }}{{\sum\nolimits_{i = 1}^K {{\text{tr}}\left( {{{\mathbf{Q}}_{il}}} \right)\bar \beta _i^{\alpha} } }},k = 1, \ldots K, l = 1, \ldots L,
\end{align}
where the channel inversion rate $\alpha = -1$ is chosen to enhance the SE performance of poor UEs and ${{\bar \beta }_k}$ denotes the global statistical channel given as
\begin{align}
{{\bar \beta }_k} = \frac{{\sum\nolimits_{l = 1}^L {{\beta _{kl}}} }}{L}.
\end{align}
Note that max-min and max-sum SE power control methods can also be applied to the case, which has been investigated in \cite{9849114,bjornson2017massive}. However, the sum SE gain of power control schemes on the high-density CF massive MIMO system is limited, because the global statistical channel difference of each UE is significantly reduced with the increase of the number of APs \cite{9849114}.

\section{Rate-Splitting Assisted CF Massive MIMO}\label{RS research}

In this section, we focus on the performance of the RS-assisted CF massive MIMO systems with asynchronous reception. By treating the private messages as noise, we first derive the closed-form SE performance expressions for the common messages. After decoding the common messages, the remaining private messages are decoded by following the same steps as in Section \ref{se:performance}. Besides, we propose a binary search-based method to find the optimal power splitting factor to maximize the downlink sum SE performance. Moreover, we design the robust precoding for the common message to mitigate the impact caused by asynchronization.

\subsection{Coherent Downlink Transmission}

Adopting the coherent joint transmission in RS-assisted CF massive MIMO systems, the transmitted signal can be divided into the common and the private parts as \eqref{data_RS}. Then, the received signal by UE $k$ at the $n$th time instant is expressed as
\begin{align}
  {y_k^\text{c,co}}\left[ n \right] &= \sum\limits_{l = 1}^L {{\mathbf{g}}_{kl}^{\text{H}}\left[ n \right]{{\mathbf{x}}_l}\left[ n \right]}  + {w_k}\left[ n \right] \notag \\
   &= \sqrt {{p_{{\text{dc}}}}} \sum\limits_{l = 1}^L {{\sqrt{\eta_l}\mathbf{g}}_{kl}^{\text{H}}\left[ n \right]{{\mathbf{v}}_{{\text{c}},l}}{s_{\text{c}}}\left[ n \right]}  + \sqrt {{p_{{\text{dp}}}}} \sum\limits_{i = 1}^K {\sum\limits_{l = 1}^L {\sqrt {{\mu _l}} {\mathbf{g}}_{kl}^{\text{H}}\left[ n \right]{{\mathbf{v}}_{il}}{s_i}\left[ n \right]} }  + {w_k}\left[ n \right] ,
\end{align}
where ${w_k}\left[ n \right]$ is the receiver noise.
Having in mind that the channels of each UE in the high antenna density regime tend to be asymptotically orthogonal \cite{7434643,8379438}, we assume that ${{\mathbf{v}}_{{\text{c}},l}}$ can be written as a linear sum of the private precoding as ${{\mathbf{v}}_{{\text{c}},l}} = \sum\nolimits_{i = 1}^K {{a_{il}}{{\mathbf{v}}_{il}}} $. It is worth noting that one simple low-complexity common precoding is given by $a_{il} = 1 , \forall i,l$, which is applied in our analysis as a benchmark scheme.
\begin{thm}\label{thm3}
For the DU-MR private precoding ${{\mathbf{v}}_{kl}} = {\theta _{kl}}{{{\mathbf{\hat h}}}_{kl}}\left[ \lambda  \right]$ and the simple common precoding ${{\mathbf{v}}_{{\mathrm{c}},l}} = \sum\nolimits_{i = 1}^K {{{\mathbf{v}}_{il}}}$, a lower bound on the achievable rate of common message of UE $k$ is
\begin{align}
{\mathrm{SE}}_k^{{\mathrm{c,co,du}}} = \frac{1}{{{\tau _c}}}\sum\limits_{n = \lambda }^{{\tau _c}} {{{\log }_2}\left( {1 + {\mathrm{SINR}}_k^{{\mathrm{c,co,du}}}\left[ n \right]} \right)} ,
\end{align}
with
\begin{align}\label{SINRc}
{\mathrm{SINR}}_k^{\mathrm{c,co,du}}\left[ n \right] = \frac{{{{p_{{\mathrm{dc}}}}\eta _n^{{\mathrm{ap}}}\eta _n^{{\mathrm{ue}}}{{\left| {\sum\limits_{l = 1}^L {\sqrt {{\eta _l}} \sum\limits_{i \in {\mathcal{P}_k}}^K {{a_{il}}{\mathrm{tr}}\left( {{{{\mathbf{\bar Q}}}_{kil}}} \right)} } } \right|}^2}}}}{{{{p_{{\mathrm{dc}}}}\Gamma _k^{{\mathrm{c,co,du}}}\left[ n \right]} + {p_{{\mathrm{dp}}}}{\Xi _k^{{\mathrm{p,co}},{\mathrm{du}}}\left[ n \right]} + \sigma _{\mathrm{d}}^2}} ,
\end{align}
where
\begin{align}
\Gamma _k^{{\mathrm{c,co,du}}}\left[ n \right] &= \left( {1 - \eta _n^{{\mathrm{ap}}}} \right)\sum\limits_{l = 1}^L {{\eta _l}\sum\limits_{i \in {\mathcal{P}_k}}^K {\sum\limits_{j \in {\mathcal{P}_i}}^K {a_{il}^*{a_{jl}}{\mathrm{tr}}\left( {{{{\mathbf{\bar Q}}}_{kil}}} \right)} } {\mathrm{tr}}\left( {{{{\mathbf{\bar Q}}}_{kjl}}} \right)}  \notag\\
&+ \sum\limits_{l = 1}^L {{\eta _l}\sum\limits_{i = 1}^K {\sum\limits_{j \in {\mathcal{P}_i}}^K {a_{il}^*{a_{jl}}{\mathrm{tr}}\left( {{{{\mathbf{\bar Q}}}_{ijl}}{{\mathbf{R}}_{kl}}} \right)} } }  + \eta _n^{{\mathrm{ap}}}\left( {1 - \eta _n^{{\mathrm{ue}}}} \right){\left| {\sum\limits_{l = 1}^L {\sqrt {{\eta _l}} \sum\limits_{i \in {\mathcal{P}_k}}^K {{a_{il}}{\mathrm{tr}}\left( {{{{\mathbf{\bar Q}}}_{kil}}} \right)} } } \right|^2} .
\end{align}
\end{thm}
\begin{IEEEproof}
See Appendix B.
\end{IEEEproof}
\begin{cor}
The same operation as in Theorem \ref{appro} for \eqref{SINRc}, its approximate SINR expression with $LN \to \infty$ is obtained as $1/\left( {1/\left(L\eta _n^{{\mathrm{ap}}}\eta _n^{{\mathrm{ue}}}\right) + \left( {L - 1} \right)/\left(L\eta _n^{{\mathrm{ue}}}\right) + b} \right)$, where $b$ is a constant. It is obvious that the decoding of common messages is weakened by the oscillator phase especially the aspect of UE, which leads to a negative impact on the SE performance of RS.
\end{cor}
Furthermore, we investigate the SE performance of the common message by adopting the DF-MR private precoding ${{\mathbf{v}}_{kl}} = {{{\mathbf{\hat h}}}_{kl}}\left[ \lambda  \right]$ and the simple common precoding. Following the similar steps as in Theorem \ref{thm3}, we obtain
\begin{align}\label{SINR_ccodf}
{\text{SINR}}_k^{{\text{c}},{\text{co}},{\text{df}}}\left[ n \right] = \frac{{{p_{{\text{dc}}}}\eta _n^{{\mathrm{ap}}}\eta _n^{{\mathrm{ue}}}{{\left| {\sum\limits_{l = 1}^L {\sqrt {{\eta _l}} \sum\limits_{i \in {\mathcal{P}_k}}^K {{a_{il}}\theta _{il}^*{\text{tr}}\left( {{{{\mathbf{\bar Q}}}_{kil}}} \right)} } } \right|}^2}}}{{{p_{{\text{dc}}}}\Gamma _k^{{\text{c}},{\text{co}},{\text{df}}}\left[ n \right] + {p_{{\text{dp}}}}{\Xi _k^{{\text{p,co}},{\text{df}}}\left[ n \right]} + {\sigma_\text{d} ^2}}} ,
\end{align}
where
\begin{align}
\Gamma _k^{{\text{c}},{\text{co}},{\text{df}}}\left[ n \right] &= \left( {1 - \eta _n^{{\mathrm{ap}}}} \right)\sum\limits_{l = 1}^L {{\eta _l}\sum\limits_{i \in {\mathcal{P}_k}}^K {\sum\limits_{j \in {\mathcal{P}_i}}^K {a_{il}^*{a_{jl}}{\theta _{il}}\theta _{jl}^{\text{*}}{\text{tr}}\left( {{{{\mathbf{\bar Q}}}_{kil}}} \right)} } {\text{tr}}\left( {{{{\mathbf{\bar Q}}}_{kjl}}} \right)}  \notag\\
&\!\!\!\!\!\!\!\!\!\!\!\!+\! \sum\limits_{l = 1}^L {{\eta _l}\sum\limits_{i = 1}^K {\sum\limits_{j \in {\mathcal{P}_i}}^K {a_{il}^*{a_{jl}}{\theta _{il}}\theta _{jl}^{\text{*}}{\text{tr}}\left( {{{{\mathbf{\bar Q}}}_{ijl}}{{\mathbf{R}}_{kl}}} \right)} } }  + \eta _n^{{\mathrm{ap}}}\left( {1 - \eta _n^{{\mathrm{ue}}}} \right){\left| {\sum\limits_{l = 1}^L {\sqrt {{\eta _l}} \sum\limits_{i \in {\mathcal{P}_k}}^K {{a_{il}}\theta _{il}^*{\text{tr}}\left( {{{{\mathbf{\bar Q}}}_{kil}}} \right)} } } \right|^2} .
\end{align}
From \eqref{SINR_ccodf}, we obtain that the delay phase effect on common messages is related to both AP and UE, specifically different physical positions of APs and sharing pilots among partial UEs.

\subsection{Non-Coherent Downlink Transmission}

When applying the non-coherent joint transmission in RS-assisted CF massive MIMO systems, each AP can send different common and private data symbols than the other APs. Therefore, the transmit signal from AP $l$ at time instant $n$ is expressed as
\begin{align}\label{data_RS_non}
{{\mathbf{x}}_l}\left[ n \right] = \sqrt {{p_{{\text{dc}}}}{\eta _l}} {{\mathbf{v}}_{{\text{c}},l}}{s_{\text{c},l}}\left[ n \right] + \sqrt {{p_{{\text{dp}}}}{\mu _l}} \sum\limits_{i = 1}^K {{{\mathbf{v}}_{il}}{s_{il}}\left[ n \right]} .
\end{align}
Compared with \eqref{data_RS}, each AP no longer cooperates with each other in \eqref{data_RS_non}, which greatly reduces the requirement for synchronization.
Then, the received signal by UE $k$ at the $n$th time instant is given by
\begin{align}
  {y_k^\text{c,nc}}\left[ n \right] &= \sum\limits_{l = 1}^L {{\mathbf{g}}_{kl}^{\text{H}}\left[ n \right]{{\mathbf{x}}_l}\left[ n \right]}  + {w_k}\left[ n \right] \notag \\
   &= \sqrt {{p_{{\text{dc}}}}} \sum\limits_{l = 1}^L {{\sqrt{\eta_l}\mathbf{g}}_{kl}^{\text{H}}\left[ n \right]{{\mathbf{v}}_{{\text{c}},l}}{s_{\text{c},l}}\left[ n \right]}  + \sqrt {{p_{{\text{dp}}}}} \sum\limits_{i = 1}^K {\sum\limits_{l = 1}^L {\sqrt {{\mu _l}} {\mathbf{g}}_{kl}^{\text{H}}\left[ n \right]{{\mathbf{v}}_{il}}{s_{il}}\left[ n \right]} }  + {w_k}\left[ n \right] .
\end{align}
\begin{thm}\label{thm4}
For the DU-MR precoding ${{\mathbf{v}}_{kl}} = {\theta _{kl}}{{{\mathbf{\hat h}}}_{kl}}\left[ \lambda  \right]$ and the simple common precoding, then the achievable rate of common message of UE $k$ is lower bounded by
\begin{align}
{\mathrm{SE}}_k^{{\mathrm{c,nc,du}}} = \frac{1}{{{\tau _c}}}\sum\limits_{n = \lambda }^{{\tau _c}} {{{\log }_2}\left( {1 + {\mathrm{SINR}}_k^{{\mathrm{c,nc,du}}}\left[ n \right]} \right)} ,
\end{align}
with
\begin{align}
{\mathrm{SINR}}_k^{{\mathrm{c,nc,du}}}\left[ n \right] = \frac{{{p_{{\mathrm{dc}}}}\eta _n^{{\mathrm{ap}}}\eta _n^{{\mathrm{ue}}}\sum\limits_{l = 1}^L {{\eta _l}} {{\left| {\sum\limits_{i \in {\mathcal{P}_k}}^K {{a_{il}}{\mathrm{tr}}\left( {{{{\mathbf{\bar Q}}}_{kil}}} \right)} } \right|}^2}}}{{{p_{{\mathrm{dc}}}}\Gamma _k^{{\mathrm{c}},{\mathrm{nc}},{\mathrm{du}}}\left[ n \right] + {p_{{\mathrm{dp}}}}\Xi _k^{{\mathrm{p,nc}},{\mathrm{du}}}\left[ n \right] + {\sigma_\mathrm{d} ^2}}} ,
\end{align}
where
\begin{align}
\Gamma _k^{{\mathrm{c}},{\mathrm{nc}},{\mathrm{du}}}\left[ n \right] = \sum\limits_{l = 1}^L {{\eta _l}\sum\limits_{i = 1}^K {\sum\limits_{j \in {\mathcal{P}_i}}^K {a_{il}^*{a_{jl}}{\mathrm{tr}}\left( {{{{\mathbf{\bar Q}}}_{ijl}}{{\mathbf{R}}_{kl}}} \right)} } }  + \left( {1 - \eta _n^{{\mathrm{ap}}}\eta _n^{{\mathrm{ue}}}} \right)\sum\limits_{l = 1}^L {{\eta _l}} {\left| {\sum\limits_{i \in {\mathcal{P}_k}}^K {{a_{il}}{\mathrm{tr}}\left( {{{{\mathbf{\bar Q}}}_{kil}}} \right)} } \right|^2} .
\end{align}
\end{thm}
\begin{IEEEproof}
Follow the similar steps as those for proving Theorems \ref{thm2} and \ref{thm3}, but treating the private messages as interference.
\end{IEEEproof}
Moreover, we study the SE performance of the common message by using DF-MR private precoding ${{\mathbf{v}}_{kl}} = {{{\mathbf{\hat h}}}_{kl}}\left[ \lambda  \right]$ and the simple common precoding ${{\mathbf{v}}_{{\text{c}},l}} = \sum\nolimits_{i = 1}^K {{{\mathbf{v}}_{il}}} $. Following the similar steps in Theorem \ref{thm4}, the SINR of the common message is derived as
\begin{align}
{\text{SINR}}_k^{{\text{c,nc,df}}}\left[ n \right] = \frac{{{p_{{\text{dc}}}}\eta _n^{{\mathrm{ap}}}\eta _n^{{\mathrm{ue}}}\sum\limits_{l = 1}^L {{\eta _l}} {{\left| {\sum\limits_{i \in {\mathcal{P}_k}}^K {{a_{il}}\theta _{il}^*{\text{tr}}\left( {{{{\mathbf{\bar Q}}}_{kil}}} \right)} } \right|}^2}}}{{{p_{{\text{dc}}}}\Gamma _k^{{\text{c}},{\text{nc}},{\text{df}}}\left[ n \right] + {p_{{\text{dp}}}}\Xi _k^{{\mathrm{p,nc}},{\mathrm{df}}}\left[ n \right] + {\sigma_\text{d} ^2}}} ,
\end{align}
where
\begin{align}\label{common_phase}
\Gamma _k^{{\text{c}},{\text{nc}},{\text{df}}}\left[ n \right] \!=\! \sum\limits_{l = 1}^L {{\eta _l}\sum\limits_{i = 1}^K {\sum\limits_{j \in {\mathcal{P}_i}}^K {a_{il}^*{a_{jl}}{\theta _{il}}\theta _{jl}^{\text{*}}{\text{tr}}\left( {{{{\mathbf{\bar Q}}}_{ijl}}{{\mathbf{R}}_{kl}}} \right)} } }  \!+\! \left( {1 \!-\! \eta _n^{{\mathrm{ap}}}\eta _n^{{\mathrm{ue}}}} \right)\sum\limits_{l = 1}^L {{\eta _l}} {\left| {\sum\limits_{i \in {\mathcal{P}_k}}^K {{a_{il}}\theta _{il}^*{\text{tr}}\left( {{{{\mathbf{\bar Q}}}_{kil}}} \right)} } \right|^2} .
\end{align}
\begin{rem}
It is worth noting that the common messages are decoded by all the UEs, which is different from the private messages only decoded by the corresponding UE. Therefore, when adopting the non-coherent transmission and the delay phase is unknown, the SE of the private message is not affected by the delay phase as stated in Remark \ref{sinr_nc}. However, the SE of the common message is still affected by the delay phase as shown in \eqref{common_phase}.
\end{rem}
\subsection{Power Splitting Factor to Maximize the Downlink Sum SE}

After the common messages have been removed from the received signal using SIC, the private messages can be decoded from the remained signal as shown in Section \ref{se:performance}.
Finally, we can obtain the downlink sum SE as
\begin{align}
{\text{SSE}} = {\text{S}}{{\text{E}}^{\text{c}}} + \sum\limits_{k = 1}^K {{\text{SE}}_k^{\text{p}}} .
\end{align}
%where
%\begin{align}
%  {\text{S}}{{\text{E}}^{\text{c}}} &= \frac{1}{{{\tau _c}}}\sum\limits_{n = {\tau _p} + 1}^{{\tau _c}} {{{\log }_2}\left( {1 + {{\min }_k}\left\{ {{\text{SINR}}_k^{\text{c}}\left[ n \right]} \right\}} \right)}  \hfill \\
%  {\text{SE}}_k^{\text{p}} &= \frac{1}{{{\tau _c}}}\sum\limits_{n = {\tau _p} + 1}^{{\tau _c}} {{{\log }_2}\left( {1 + {\text{SINR}}_k^{\text{p}}\left[ n \right]} \right)}  \hfill \\
%\end{align}

\begin{rem}
It is clear that with the increasing of the power split factor $\rho$, the SE of the common and private messages increases and decreases, respectively. Therefore, the downlink sum SE is a monotonic function and there is an optimal $\rho$ to maximize it\footnote{The optimal power splitting factor maximizing the sum SE varies with the simulation parameters, e.g., the SNR, the number of APs and UEs \cite{7470942}. Therefore, it is significant to adjust the power splitting factor for unleashing the potential of RS technology.}. The optimal $\rho$ can be found via a simple binary search method as Algorithm \ref{binary_search}.
\end{rem}

\renewcommand{\algorithmicrequire}{\textbf{Initialization:}}
\renewcommand{\algorithmicensure}{\textbf{Output:}}
\begin{algorithm}[tb]
\caption{The Proposed Binary Search for Optimal Power Allocation}
\label{binary_search}
\begin{algorithmic}[1]
\Require
Choose the initial values $\rho_\text{min} = 0$ and $\rho_\text{max} = 1$; Chose a tolerance $\varepsilon > 0$ and an increase $0 < \Delta \rho \ll \varepsilon $;
\Ensure
the power splitting factor $\rho$;
\State ${\text{SS}}{{\text{E}}_{\max }} = \text{SSE}\left( \rho^* \right) = {\text{  max}}\left\{ {\left[ {{\text{SSE}}\left( {{\rho _{\min }}} \right),{\text{SSE}}\left( {{\rho _{\max }}} \right)} \right]} \right\}$ and $\rho = \rho^*$;
\While{$\rho_\text{max}$ $-$ $\rho_\text{min}$ $>$ $\varepsilon$}
\State Set $\rho_\text{next} = \left(\rho_\text{max} + \rho_\text{min}\right)/2$;
\State $\text{SSE}_\text{next} = \text{SSE}\left(\rho_\text{next}\right)$ and $\text{SSE}_{\Delta} = \text{SSE}\left(\rho_\text{next} + \Delta \rho\right)$;
\State If $\text{SSE}_{\Delta} > \text{SSE}_\text{next}$, then set $\rho_\text{min} = \rho_\text{next}$, else set $\rho_\text{max} = \rho_\text{next}$;
\State If $\text{SSE}_\text{next} > \text{SSE}_\text{max}$, then set $\text{SSE}_\text{max} = \text{SSE}_\text{next}$ and $\rho \triangleq \rho_\text{next}$;
\EndWhile
\end{algorithmic}
\end{algorithm}

\subsection{Robust Precoding Design for Common Message}

In the downlink, given realizations of the large-scale fading, we design the precoding weights for the common messages\footnote{Here, we focus on the precoding of the common message, since the precoding technology for the private message in CF massive MIMO is already well established, including MR, zero-forcing, and MMSE \cite{Ngo2017Cell,bjornson2019making}.} $a_{il}, i = 1, \ldots ,K, l = 1, \ldots ,L$, that maximize the minimum of the downlink common SE of all the UEs, under the power constraint. At the optimum point, all the users should achieve the same SE, then we have following max-min optimization problem:
\begin{align}\label{maxmin1}
\begin{gathered}
  \mathop {\max }\limits_{\left\{ {{a}_{kl}} \right\}}\  \mathop {\min }\limits_{k = 1, \ldots ,K}\  {\text{SE}}_k^{\text{c}}\left[ n \right] \hfill \\
  \ \ \ \ \ \ \ {\text{s}}.{\text{t}}.\ \ \ \ \ \ \mathbb{E}\left\{ {{{\left\| {{{\mathbf{v}}_{{\text{c}},l}}} \right\|}^2}} \right\} \leqslant 1,\forall l , \hfill \\
  \ \ \ \ \ \ \ \ \ \ \ \ \ \ \ \ \ {{a}_{kl}} \geqslant 0, \forall k ,l , \hfill \\
\end{gathered}
\end{align}
where ${\text{SE}}_k^{\text{c}}\left[ n \right]$ is given by \eqref{SINRc}. After some straight-forward transformation of \eqref{SINRc}, the problem in \eqref{maxmin1} is equivalent to\footnote{Compared with \cite{9737523}, although the same motivation is adopted to design the precoding for the common messages, our design takes into account the residual interference for the common message precoder while the former did not.}
\begin{align}\label{maxmin2}
\begin{gathered}
  \mathop {\max }\limits_{\left\{ {\mathbf{a}} \right\}} \mathop {\min }\limits_{k = 1, \ldots ,K} \frac{{{p_{{\text{dc}}}}\eta _n^{{\text{ap}}}\eta _n^{{\text{ue}}}{{\left| {{{\mathbf{a}}^{\text{H}}}{{\mathbf{b}}_k}} \right|}^2}}}{{{p_{{\text{dc}}}}\left( {1 \!-\! \eta _n^{{\text{ap}}}} \right){{\mathbf{a}}^{\text{H}}}{{\mathbf{H}}_k}{\mathbf{a}} \!+\! {p_{{\text{dc}}}}{{\mathbf{a}}^{\text{H}}}{{\mathbf{M}}_k}{\mathbf{a}} \!+\! {p_{{\text{dc}}}}\eta _n^{{\text{ap}}}\left( {1 \!-\! \eta _n^{{\text{ue}}}} \right){{\left| {{{\mathbf{a}}^{\text{H}}}{{\mathbf{b}}_k}} \right|}^2} \!+\! {p_{{\text{dp}}}}\Xi _k^{{\text{co}},{\text{du}}}\left[ n \right] \!+\! \sigma _{\text{d}}^2}} \hfill \\
  \ \ \ \ \ \ \ \ {\text{s}}.{\text{t}}.\ \ \ \left\| {{\mathbf{\Theta }}_l^{\frac{1}{2}}{{\mathbf{a}}_l}} \right\| \leqslant 1,\forall l , \hfill \\
  \ \ \ \ \ \ \ \ \ \ \ \ \ \ \ a_{kl},  \geqslant 0 , \forall k,l, \hfill \\
\end{gathered}
\end{align}
with
\begin{align}
{\mathbf{a}} &= {[\mathbf{a}_{1}^{\mathrm{T}}, \ldots ,\mathbf{a}_{L}^{\mathrm{T}}]^{\mathrm{T}}} \in {\mathbb{C}^{KL}} , \\
{{\mathbf{b}}_k} &= {[{b_{k11}}, \ldots ,{b_{kK1}}, \ldots ,{b_{k1L}}, \ldots ,{b_{kKL}}]^{\mathrm{T}}} \in {\mathbb{C}^{KL}} ,  \\
{{\mathbf{{ H}}}_k} &= {\mathrm{diag}}\left( {{{\mathbf{{ H}}}_{k1}}, \ldots ,{{\mathbf{{ H}}}_{kL}}} \right) \in {\mathbb{C}^{KL \times KL}} ,   \\
{{\mathbf{{ M}}}_k} &= {\mathrm{diag}}\left( {{{\mathbf{M}}_{k1}}, \ldots ,{{\mathbf{M}}_{kL}}} \right) \in {\mathbb{C}^{KL \times KL}} ,   \\
{{\mathbf{\Theta }}_l} &= \left[ {\begin{array}{*{20}{c}}
  {{b_{11l}}}& \cdots &{{b_{1Kl}}} \\
   \vdots & \ddots & \vdots  \\
  {{b_{K1l}}}& \cdots &{{b_{KKl}}}
\end{array}} \right]  \in {\mathbb{C}^{K \times K}} ,
\end{align}
where
\begin{align}
{\mathbf{a}_l} &= {[{a}_{1l}, \ldots ,{a}_{K1}]^{\mathrm{T}}} \in {\mathbb{C}^{K}} ,   \\
{b_{kil}} &= \left\{ {\begin{array}{*{20}{c}}
  {{\mathrm{tr}}\left( {{{{\mathbf{\bar Q}}}_{kil}}} \right),i \in {\mathcal{P}_k}}  \\
  {0,i \notin {\mathcal{P}_k}}
\end{array}} \right. , \\
{{\mathbf{{ H}}}_{kl}} &= \left[ {\begin{array}{*{20}{c}}
  {{h_{k11}}}& \cdots &{{h_{k1K}}} \\
   \vdots & \ddots & \vdots  \\
  {{h_{kK1}}}& \cdots &{{h_{kKK}}}
\end{array}} \right] \mathrm{with}\  {h_{kij}} = \left\{ {\begin{array}{*{20}{c}}
  {{\mathrm{tr}}\left( {{{{\mathbf{\bar Q}}}_{kil}}} \right){\mathrm{tr}}\left( {{{{\mathbf{\bar Q}}}_{kjl}}} \right),i,j \in {\mathcal{P}_k}} \\
  {0,i \notin {\mathcal{P}_k}/j \notin {\mathcal{P}_k}}
\end{array}} \right. ,  \\
{{\mathbf{M}}_{kl}} &= \left[ {\begin{array}{*{20}{c}}
  {{m_{k11}}}& \cdots &{{m_{k1K}}} \\
   \vdots & \ddots & \vdots  \\
  {{m_{kK1}}}& \cdots &{{m_{kKK}}}
\end{array}} \right] \mathrm{with}\  {m_{kij}} = \left\{ {\begin{array}{*{20}{c}}
  {{\mathrm{tr}}\left( {{{{\mathbf{\bar Q}}}_{ijl}}{{\mathbf{R}}_{kl}}} \right),j \in {\mathcal{P}_i}} \\
  {0,j \notin {\mathcal{P}_i}}
\end{array}} \right. .
\end{align}

\renewcommand{\algorithmicrequire}{\textbf{Initialization:}}
\renewcommand{\algorithmicensure}{\textbf{Output:}}
\begin{algorithm}[tb]
\caption{Bisection Algorithm for Solving \eqref{maxmin2}}
\label{bisection}
\begin{algorithmic}[1]
\Require
Initialize $t_\text{min}$ and $t_\text{max}$, where $t_\text{min}$ and $t_\text{max}$ define a range of relevant values of the objective function in \eqref{maxmin2}. Choose a tolerance $\varepsilon > 0$.
\Ensure
The percoding weight matrix $\mathbf{a}$;
\While{$t_\text{max}$ $-$ $t_\text{min}$ $>$ $\varepsilon$}
\State Set $t = \left(t_\text{max} + t_\text{min}\right)/2$. Solve the following convex feasibility program:
\begin{align}\label{bis}
\left\{ {\begin{array}{*{20}{c}}
  {\sqrt {{p_{{\text{dc}}}}\eta _n^{{\text{ap}}}\eta _n^{{\text{ue}}}} {{\mathbf{a}}^{\text{H}}}{{\mathbf{b}}_k} \geqslant \sqrt t \left\| {{{\mathbf{u}}_{n,k}}} \right\|,\forall k}, \\
  {\left\| {{\mathbf{\Theta }}_l^{\frac{1}{2}}{{\mathbf{a}}_l}} \right\| \leqslant 1,\forall l}, \\
  {a_{kl}  \geqslant 0,  \forall k,l .}
\end{array}} \right.
\end{align}
\State Besides, ${{\mathbf{u}}_{n,k}}$ is defined as
\begin{align}
{\left[ {{{\left(\! {\sqrt {{p_{{\text{dc}}}}\left( {1 \!-\! \eta _n^{{\text{ap}}}} \right)} {\mathbf{H}}_k^{\frac{1}{2}}{\mathbf{a}}} \!\right)}^{\text{T}}},{{\left(\! {\sqrt {{p_{{\text{dc}}}}} {\mathbf{M}}_k^{\frac{1}{2}}{\mathbf{a}}} \!\right)}^{\text{T}}},\sqrt {{p_{{\text{dc}}}}\eta _n^{{\text{ap}}}\left( {1 \!-\! \eta _n^{{\text{ue}}}} \right)} {{\mathbf{a}}^{\text{H}}}{{\mathbf{b}}_k},\sqrt {{p_{{\text{dp}}}}\Xi _k^{{\text{co}},{\text{du}}}\left[ n \right] \!+\! \sigma _{\text{d}}^2} } \right]^{\text{T}}} \notag
\end{align}
\State If problem \eqref{bis} is feasible, then set $t_\text{min} \triangleq t$, else set $t_\text{max} \triangleq t$.
\EndWhile
\end{algorithmic}
\end{algorithm}

With the help of \cite[Proposition 1]{Ngo2017Cell}, it can be shown that the problem in \eqref{maxmin2} is quasi-concave.
Therefore, the problem in \eqref{maxmin2} can be efficiently solved by using the bisection method, which solving a sequence of convex feasibility problems in each step. The detailed scheme is presented in Algorithm~\ref{bisection}.
The outer while-loop in Algorithm~\ref{bisection} performs a bisection search for the optimal SINR value, which halves the search space for the max-min SINR value in every iteration. Therefore, the overall algorithm converge, rapidly.

\section{Numerical Results and Discussion}\label{se:numerical}

%\begin{figure}[t]
%\centering
%\includegraphics[scale=0.55]{CDF.eps}
%\caption{CDF of the downlink sum SE for CF massive MIMO systems with DU-MR combining under coherent transmission ($L=40$, $K=8$, $N=2$, $\tau_p = 4$).} \vspace{-4mm}
%\label{CDF}
%\end{figure}
We adopt the three-slope propagation model in a simulation setup in which $L$ APs and $K$ UEs are uniformly and independently distributed within a square of size $100$ m $\times$ $100$ m\footnote{Deploying high-density APs in a small area facilitates the establishment of channel hardening and favorable propagation to ensure the accuracy of our results \cite{8379438}.} \cite{Ngo2017Cell}.
%\begin{align}
%{{\beta _{kl}}}\left[ {{\mathrm{dB}}} \right] = \left\{ {\begin{array}{*{20}{c}}
%  { - 81.2,{d_{kl}} < 10\ {\mathrm{m}}} , \\
%  { - 61.2 - 20{{\log }_{10}}\left( {\frac{{{d_{kl}}}}{{1{\mathrm{m}}}}} \right),10\ {\mathrm{m}} \leqslant {d_{kl}} < 50\ {\mathrm{m}}} , \\
%  { - 35.7 - 35{{\log }_{10}}\left( {\frac{{{d_{kl}}}}{{1{\mathrm{m}}}}} \right) + {F_{kl}},{d_{kl}} \geqslant 50\ {\mathrm{m}}},
%\end{array}} \right.
%\end{align}
%where $d_{kl}$ denotes the horizontal distance between AP $l$ and UE $k$. When the distance is longer than 50 m and the shadowing terms ${F_{kl}} \sim \mathcal{N}\left( {0,{8^2}} \right)$ are correlated as
%\begin{align}
%\mathbb{E}\left\{ {{F_{kl}}{F_{ij}}} \right\} = \frac{{{8^2}}}{2}\left( {{2^{ - {\delta _{ki}}/100{\mathrm{m}}}} + {2^{ - {\upsilon _{lj}}/100{\mathrm{m}}}}} \right),
%\end{align}
%where $\delta _{ki}$ represents the distance between the $k$th UE and $i$th UE, $\upsilon  _{lj}$ represents the distance between the $l$th AP and $j$th AP.
It is assumed that the carrier frequency is $f_c=2$ GHz, the bandwidth is $B=20$ MHz, and the noise power is $\sigma^2=-96$ dBm. In addition, the pilot and data transmission power are $p = 20$ dBm and $p_\text{d} = 23$ dBm, respectively \cite{Ngo2017Cell}. Besides, the length of a coherence block is $T = 2$ ms consisting of $\tau_c = 200$ channel uses \cite{bjornson2019making}. Moreover, the symbol duration $T_s = 10\ \mu\text{s}$ and the oscillator coefficients for all the APs and the UEs are $c_i = 1\times10^{-18}, i=k,l$ \cite{bjornson2015massive}.

\begin{figure}[t]
\begin{minipage}[t]{0.48\linewidth}	
\centering
\includegraphics[scale=0.55]{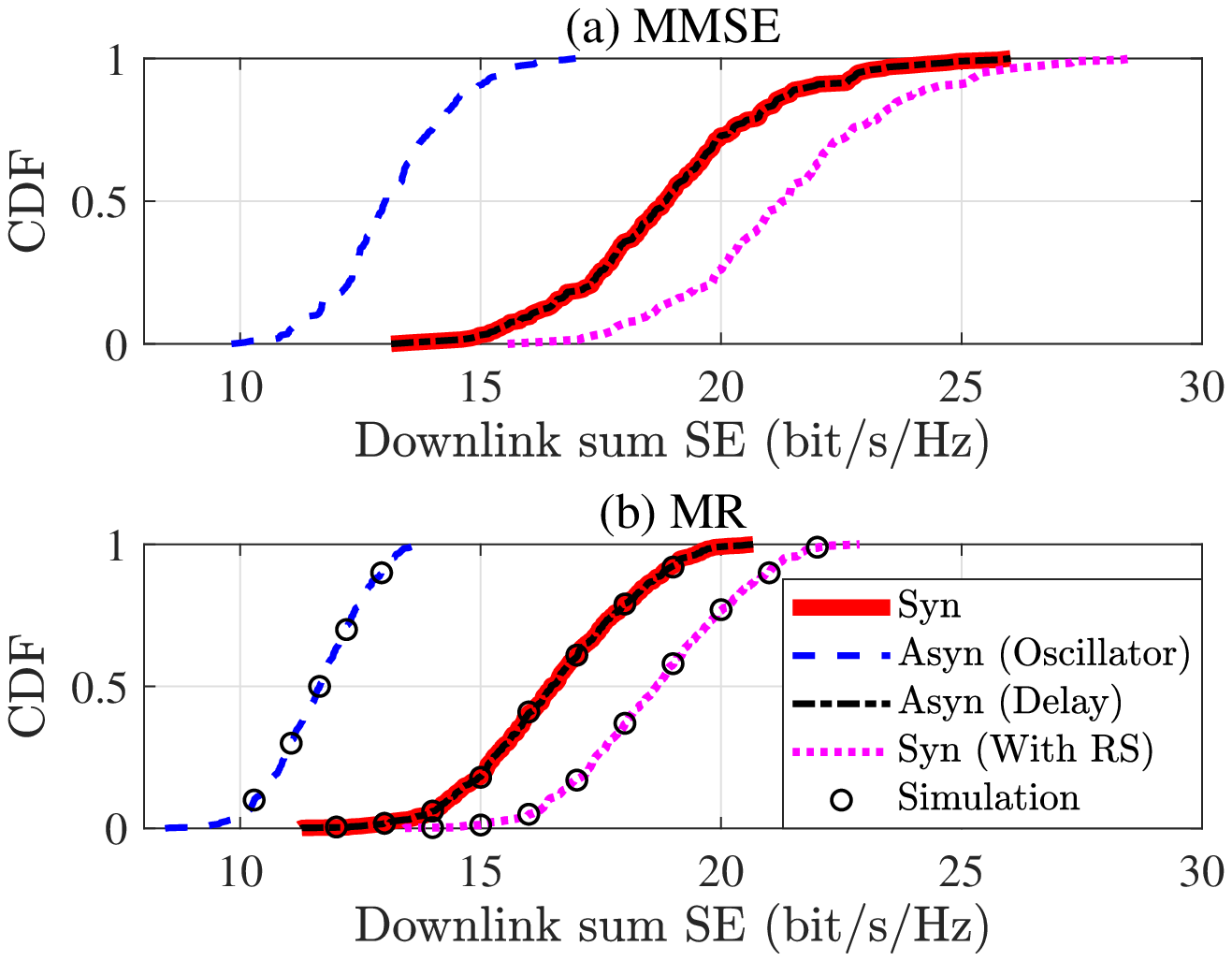}
\caption{CDF of the downlink sum SE for CF massive MIMO systems under coherent transmission ($L = 40$, $K = 8$, $N = 2$, $\tau_p = 4$). (a) DU-MMSE; (b) DU-MR.} \vspace{-4mm}
\label{CDF}
\end{minipage}
\hfill
\begin{minipage}[t]{0.48\linewidth}
\centering
\includegraphics[scale=0.55]{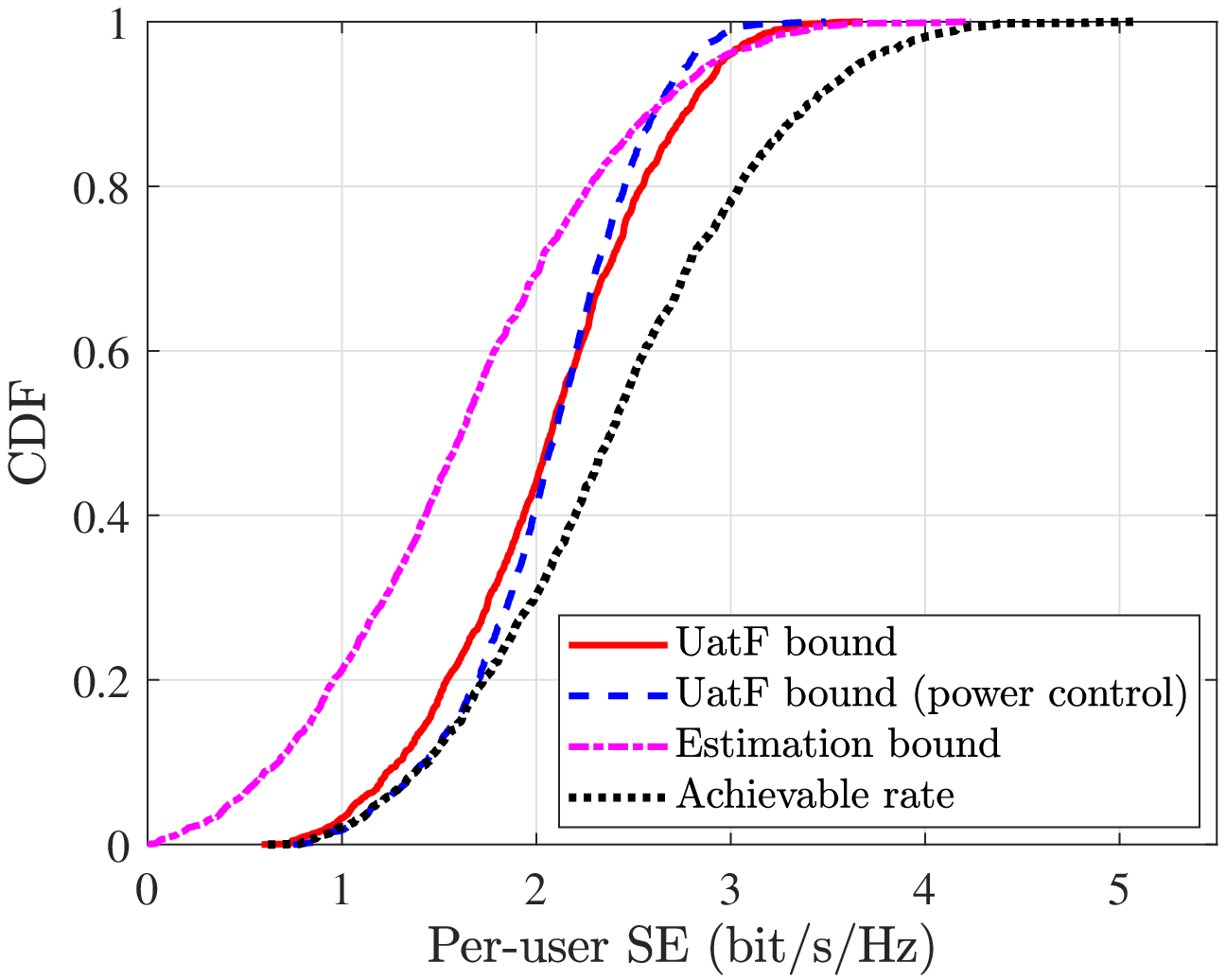}
\caption{CDF of the downlink per-user SE for synchronous CF massive MIMO systems with DU-MR combining under coherent transmission ($L=40$, $K=8$, $N=2$, $\tau_p = 4$).} \vspace{-4mm}
\label{Bound_PC}
\end{minipage}
\end{figure}

Taking into account the oscillator and delay phase respectively, Fig.~\ref{CDF} shows the CDF of the downlink sum SE for the considered CF massive MIMO under the coherent transmission with DU-MMSE and DU-MR precoding. It is clear that the DU-MMSE precoding outperforms the DU-MR precoding, due to its excellent interference cancellation capability. Moreover, the oscillator phase has a deteriorated effect on the SE performance, as the coherent data transmission gain of CF massive MIMO systems is destroyed by the asynchronous reception. Besides, since the delay phase is perfectly known and is adopted for DU-MR precoding, the coherent transmission and the SE performance is uneffected, as accurately predicted by \eqref{SINR_CF}. We also find that the application of the RS can increase the downlink sum SE by 2.2 bit/s/Hz under DU-MR precoding, due to part of the interference is broadcasted such that it is decoded and cancelled by all the UEs. Moreover, our simulation results have also confirmed the accuracy of our derived closed-form SE expressions.

Considering different capacity-bound techniques, including UatF bound \cite{bjornson2017massive}, estimation bound \cite{8379438}, and achievable rate (UEs have perfect CSI \cite{demir2021foundations}), Fig.~\ref{Bound_PC} shows the CDF of the downlink per-user SE for synchronous CF massive MIMO systems with DU-MR combining under coherent transmission. It is found that the difference in median SE performance between the UatF bound and the achievable rate is 0.3 bit/s/Hz, but the estimation bound and the achievable rate have a median SE performance difference of 0.8 bit/s/Hz. The reason is that the high antenna density ($4000 \text{ APs}/\text{km}^2$)\footnote{The CF massive MIMO system allows the high-density deployment of APs, e.g., via radio strips, which is highly practical for deploying to some hot spots, like railway stations, museums and factories \cite{9499049}.} enhances the channel hardening such that the UatF rate becomes a tight bound and the varying oscillator phase limits the use of long coherent block length. Moreover, we also find that the statistical channel cooperation power control can achieve a good SE performance gain for UEs with poor channel conditions.

\begin{figure}[t]
\begin{minipage}[t]{0.48\linewidth}	
\centering
\includegraphics[scale=0.55]{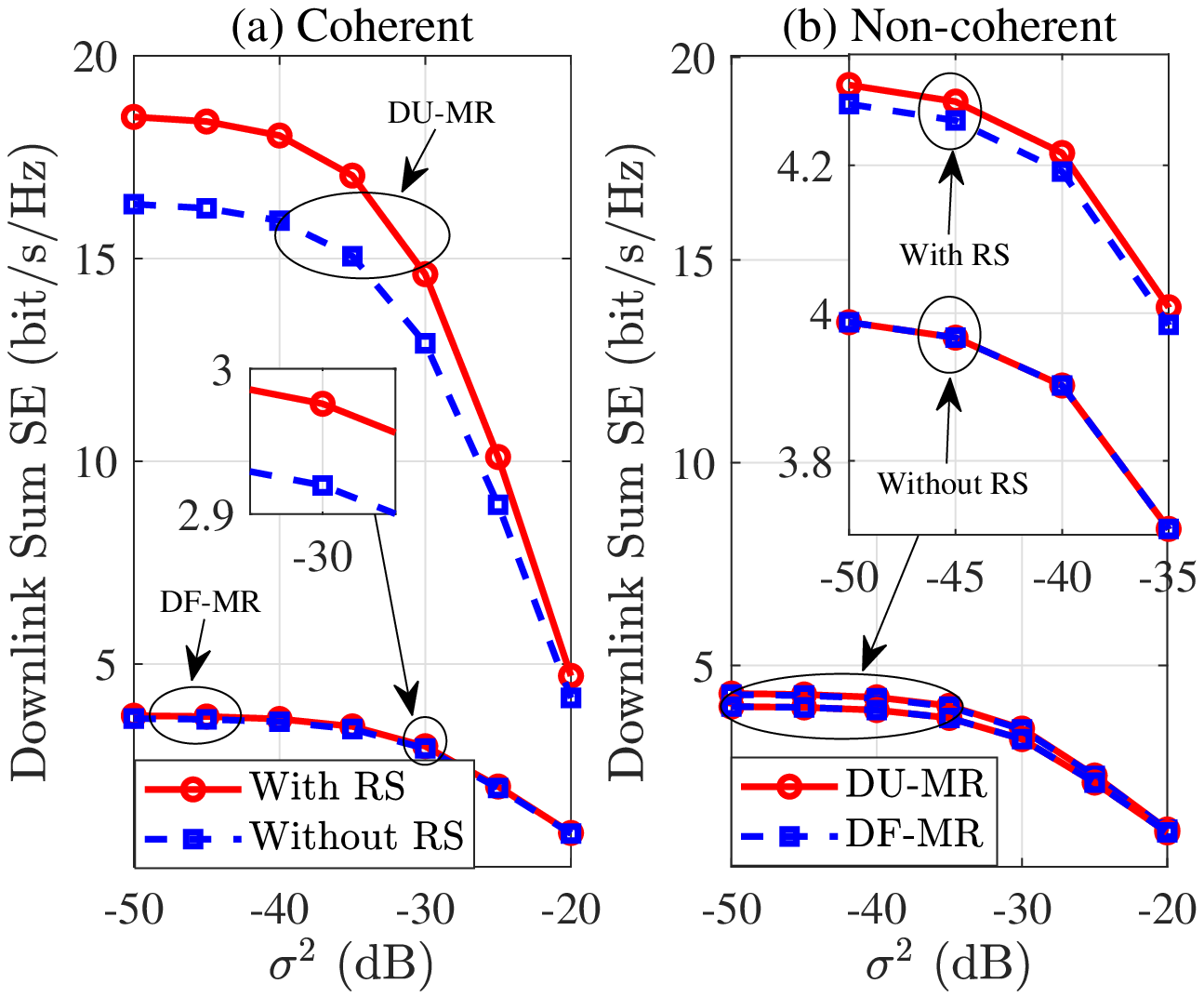}
\caption{Downlink sum SE for CF massive MIMO systems with different oscillator phase variances ($L=40$, $K=8$, $N=2$, $\tau_p = 4$, $\sigma_\text{ap}^2=\sigma_\text{ue}^2=\sigma^2$). (a) Coherent transmission; (b) Non-coherent transmission} \vspace{-4mm}
\label{coherent}
\end{minipage}
\hfill
\begin{minipage}[t]{0.48\linewidth}
\centering
\includegraphics[scale=0.55]{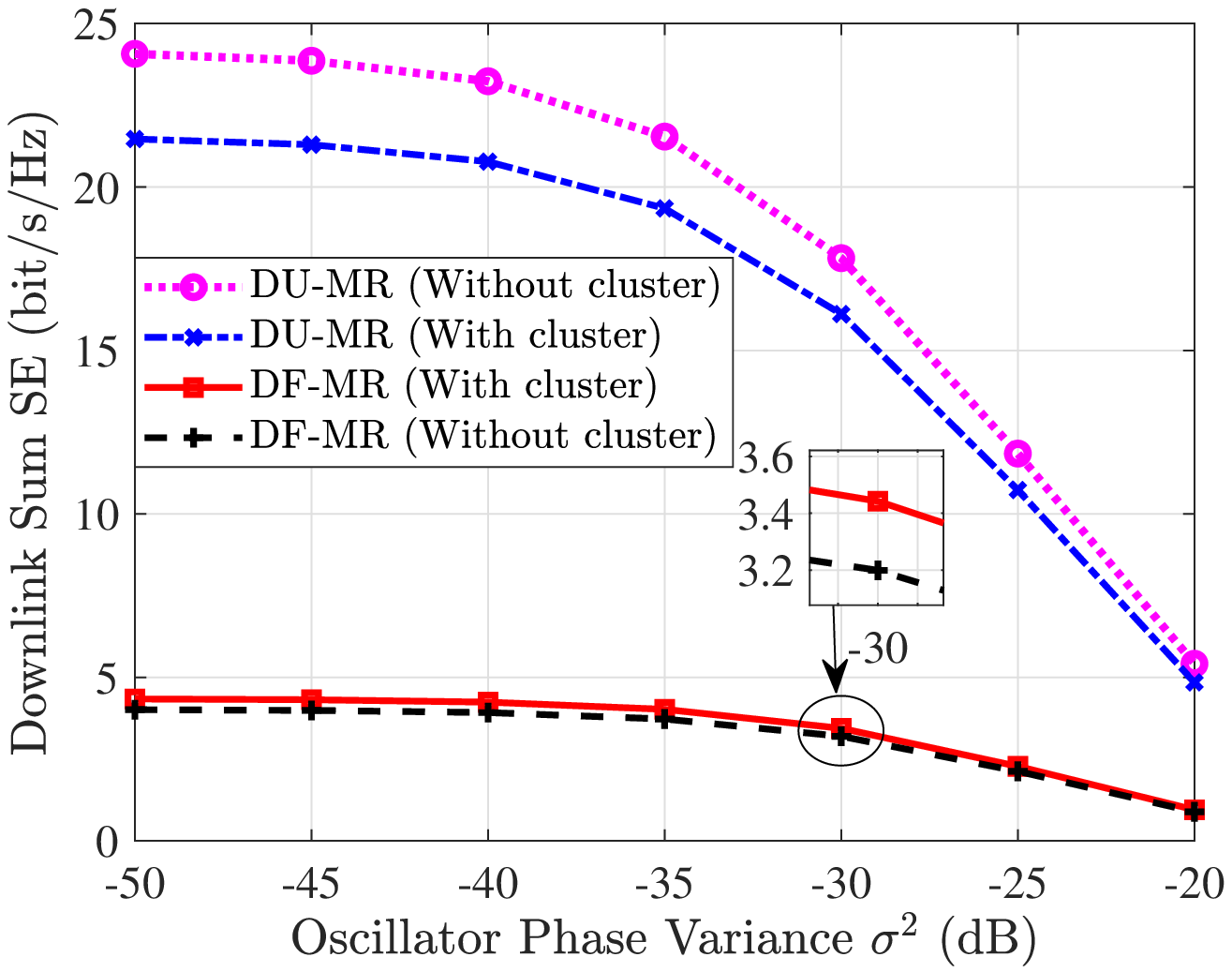}
\caption{Downlink sum SE for CF massive MIMO systems with different oscillator phase variances under coherent transmission ($L=40$, $K=8$, $N=2$, $\tau_p = K$, $\sigma_\text{ap}^2=\sigma_\text{ue}^2=\sigma^2$).} \vspace{-4mm}
\label{cluster}
\end{minipage}
\end{figure}

Fig.~\ref{coherent} compares the downlink sum SE of the CF massive MIMO system against the oscillator phase variance under the coherent and non-coherent transmissions. It is clear that the performance of the coherent transmission and DU-MR drops rapidly with the oscillator phase variance increases. The reason is that coherent transmission requires highly accurate synchronization. For example, in the case with DF-MR, the existence of the delay phase degrades the efficiency of coherent transmission. Note that the delay phase caused by the offset of heterogeneous propagation distances is often several hundred times compared to the wavelength ($\lambda$) that affects the system much more severer than the oscillator phase. Therefore, for the precoding design of the distributed architecture, it is necessary to obtain the accurate delay phase by some advanced methods such as positioning technology \cite{9650567,5958644}. In addition, we also find that when the oscillator phase variance varies from $-50$ dB to $-20$ dB, the sum SE gain of RS in the case with DU-MR precoding decreases from $2.2$ bit/s/Hz to $0.5$ bit/s/Hz. Even worse, for the case with DF-MR precoding, the sum SE gain of having RS is less more $0.1$ bit/s/Hz. This is because accurate synchronization is the key to realize the gains of RS.
For the non-coherent case in Fig.~\ref{coherent} (b) that there is no cooperation among APs, the delay phase has no impact on the CF massive MIMO with non-coherent transmission when the RS technology is not adopted. In fact, as a certain accuracy in the delay synchronization is needed for RS, the sum SE gain of RS with DU-MR precoding outperforms the case with DF-MR precoding under non-coherent transmission. Besides, by comparing Fig.~\ref{coherent} (a) and Fig.~\ref{coherent} (b), we also find that the sum SE performance of non-coherent transmission case is better than that of the coherent transmission when the delay phase is unknown. In particular, the random phase shift not only hinders effective cooperation among the APs, but also increases the multiuser interference.

Taking the user-centric concept into consideration and achieving it through dynamic cooperation clustering \cite{8761828,demir2021foundations,9849114}, Fig.~\ref{cluster} presents the downlink sum SE of CF massive MIMO systems against the oscillator phase variance under coherent transmission. In order to verify whether the dynamic cooperation clustering has an impact on the asynchronous reception error, we do not consider pilot contamination by setting $\tau_p = K$ to ensure a relatively fair comparison. It can be found that dynamic cooperation clustering does have a certain positive effect on CF massive MIMO systems when the delay phase is unaware, but it is not obvious. This is due to the fact that a small cluster size aids in overcoming the issue of geographically distributed APs introducing differences in the signal arrival time. However, when compared with the wavelength, the offset of heterogeneous propagation distances within the same cluster is still considerably large. Therefore, dynamic cooperation clustering cannot fundamentally eliminate the asynchronous reception error but can be adopted to assist existing synchronization schemes to achieve low-complexity and high-speed calibration in CF massive MIMO systems \cite{6760595}.

\begin{figure}[t]
\begin{minipage}[t]{0.48\linewidth}	
\centering
\includegraphics[scale=0.55]{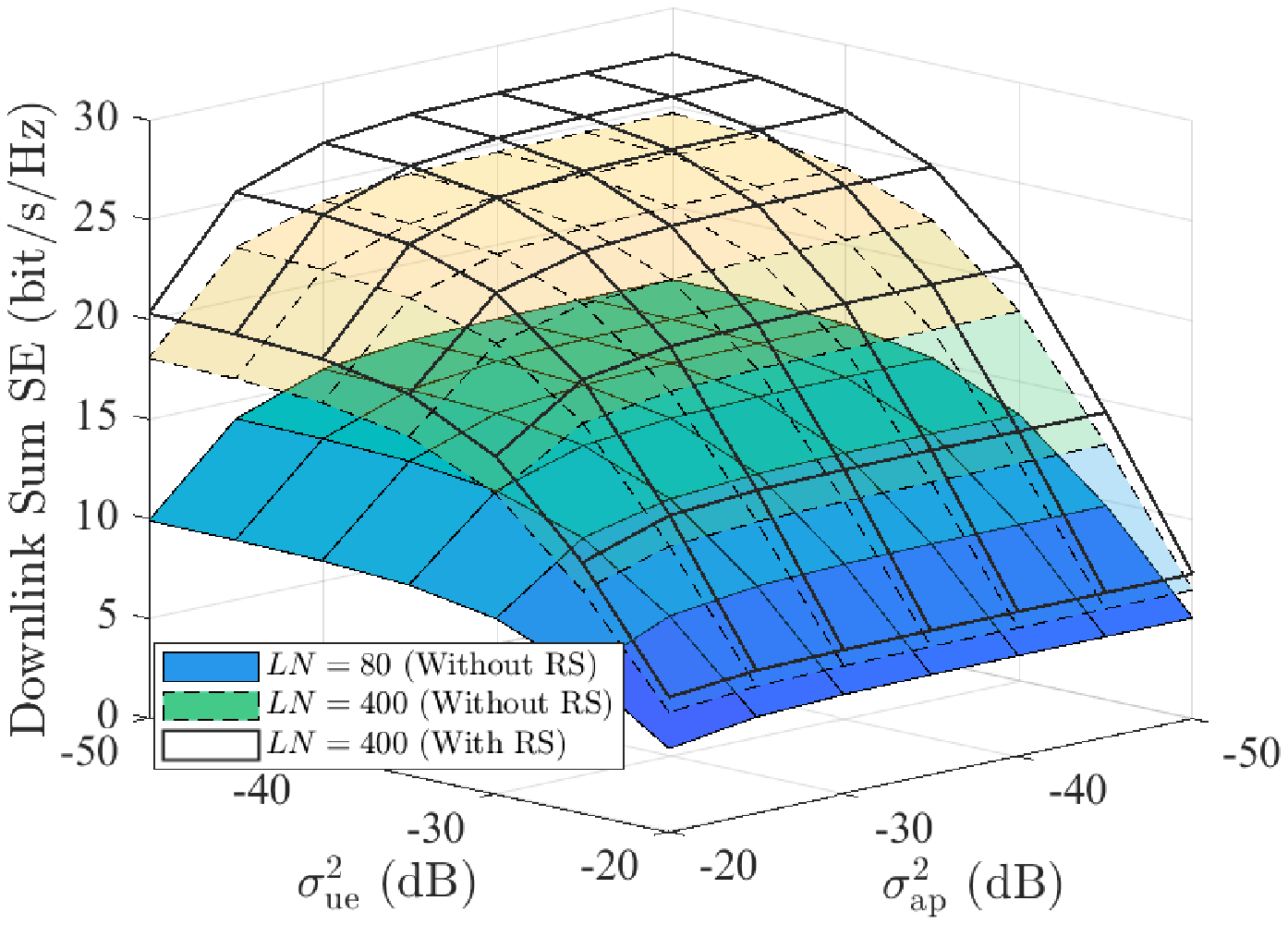}
\caption{Downlink sum SE for CF massive MIMO systems against different oscillator phase variances at AP and UE ($K=8$, $\tau_p = 4$).} \vspace{-4mm}
\label{oscillatorAPUE}
\end{minipage}
\hfill
\begin{minipage}[t]{0.48\linewidth}
\centering
\includegraphics[scale=0.55]{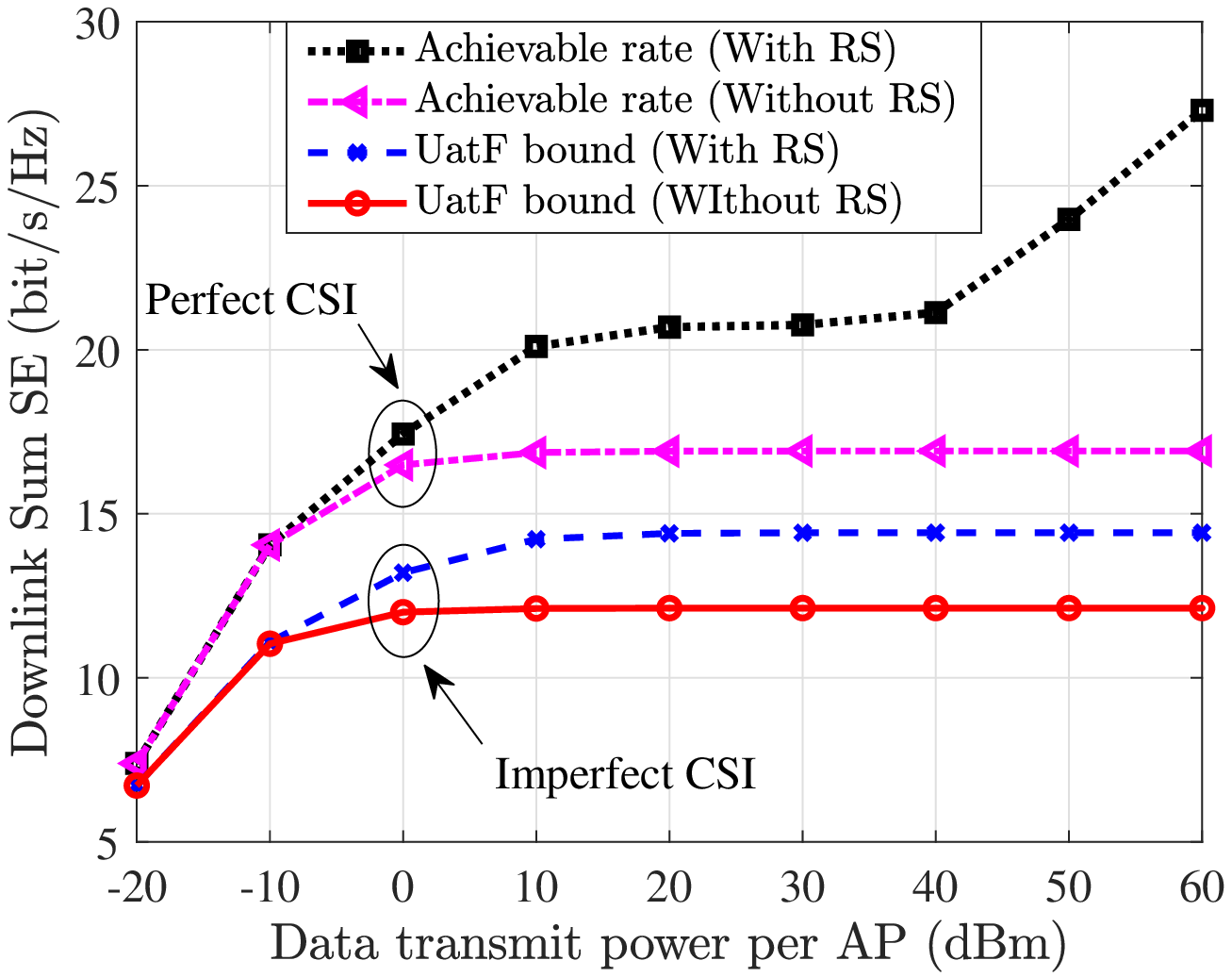}
\caption{Downlink sum SE for CF massive MIMO systems against different transmit power per AP ($L=20$, $K=4$, $N=2$, $\tau_p = 4$).} \vspace{-4mm}
\label{Pmax}
\end{minipage}
\end{figure}

The downlink sum SE for the CF massive MIMO system with the DU-MR precoding and coherent transmission is shown in Fig.~\ref{oscillatorAPUE}, as a decreasing function of the oscillator phase variances at AP and UE. We notice that the oscillator phase of the UE has a more significant negative impact on the SE performance than that of the APs. For instance, for the case $LN=80$, increasing $\sigma^2_\text{ap}$ from $-50$ dB to $-20$ dB will result in $6.4$ bit/s/Hz sum SE loss, but the same amount of increment on $\sigma^2_\text{ue}$ would cause $11.3$ bit/s/Hz sum SE loss. It is worthy noting that the negative influence becomes more pronounced as the number of antennas increases, which also can be obtained by Remark~\ref{appro}.
Moreover, it is shown that varying the number of antennas from $LN=80$ to $LN=400$ introduces a sum SE gain of $1.4$ bit/s/Hz for the case $\sigma^2_\text{ap}=-50$ dB, $\sigma^2_\text{ue}=-20$ dB, and leads to $8.1$ bit/s/Hz sum SE gain for the case $\sigma^2_\text{ap}=-20$ dB, $\sigma^2_\text{ue}=-50$ dB.
The reason is that the more antennas promise higher antenna array gains offering more degrees of freedom to compensate the negative impact caused by the oscillator phase at the AP.
Besides, due to the common messages are decoded by all the UEs, the performance gain of the RS is also more sensitive to the oscillator phase of the UEs rather than that of the APs. For example, the sum SE gain of RS for the case $\sigma^2_\text{ap}=-20$ dB, $\sigma^2_\text{ue}=-50$ dB is $2.2$ bit/s/Hz, but the sum SE gain of it for the case $\sigma^2_\text{ap}=-50$ dB, $\sigma^2_\text{ue}=-20$ dB is only 0.8 bit/s/Hz.

Taking into account the impacts of perfect and imperfect CSI, Fig.~\ref{Pmax} shows the downlink sum SE for CF massive MIMO systems against the data transmit power under coherent transmission and DU-MR private precoding. It is clear that the downlink sum SE of conventional CF massive MIMO systems (UatF bound without RS) gradually saturates as the data transmit power increases. The reason is that although the power of the desired signal increases, the power of inter-user interference due to imperfect CSI also increases. Besides, it can be observed that RS can enhance the system performance in such cases, but with diminishing returns in the high transmit power regime.
In fact, the residual interference caused by the beamforming gain uncertainty term of the UatF bound always leads to the downlink sum SE performance saturation, even after the application of RS. From Fig.~\ref{Pmax}, it is observed that the achievable rate in perfect CSI without RS also saturates due to the strong inter-user interference caused by asynchronous issues. Remarkably, RS has shown its robustness in this case, since the downlink sum SE does not saturate due to the introduction of the common message. Due to the power splitting factors for each AP are assumed to be the same, the sum achievable rate of the RS-assisted CF massive MIMO system grows slowly with the increase of the transmit power when the RS starts to work.
%Besides, when the transmit power is small and large, the achievable rate gains of RS-assisted CF massive MIMO systems mainly depend on the private and common messages, respectively.

\begin{figure}[t]
\begin{minipage}[t]{0.48\linewidth}	
\centering
\includegraphics[scale=0.55]{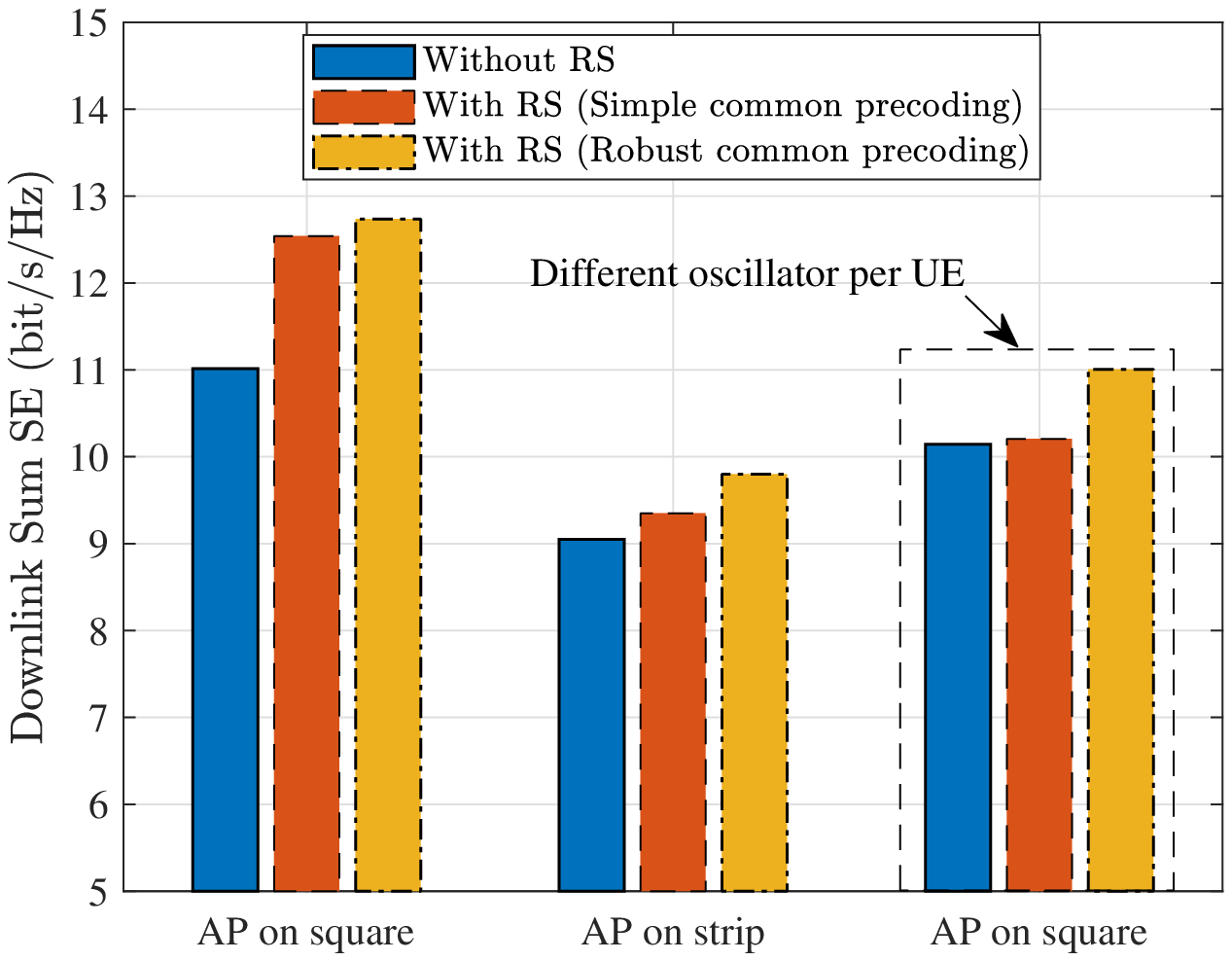}
\caption{Downlink sum SE for asynchronous CF massive MIMO systems with three signal processing schemes under different AP settings ($L=40$, $K=8$, $N=2$, $\tau_p = 4$).} \vspace{-4mm}
\label{RandLine}
\end{minipage}
\hfill
\begin{minipage}[t]{0.48\linewidth}
\centering
\includegraphics[scale=0.55]{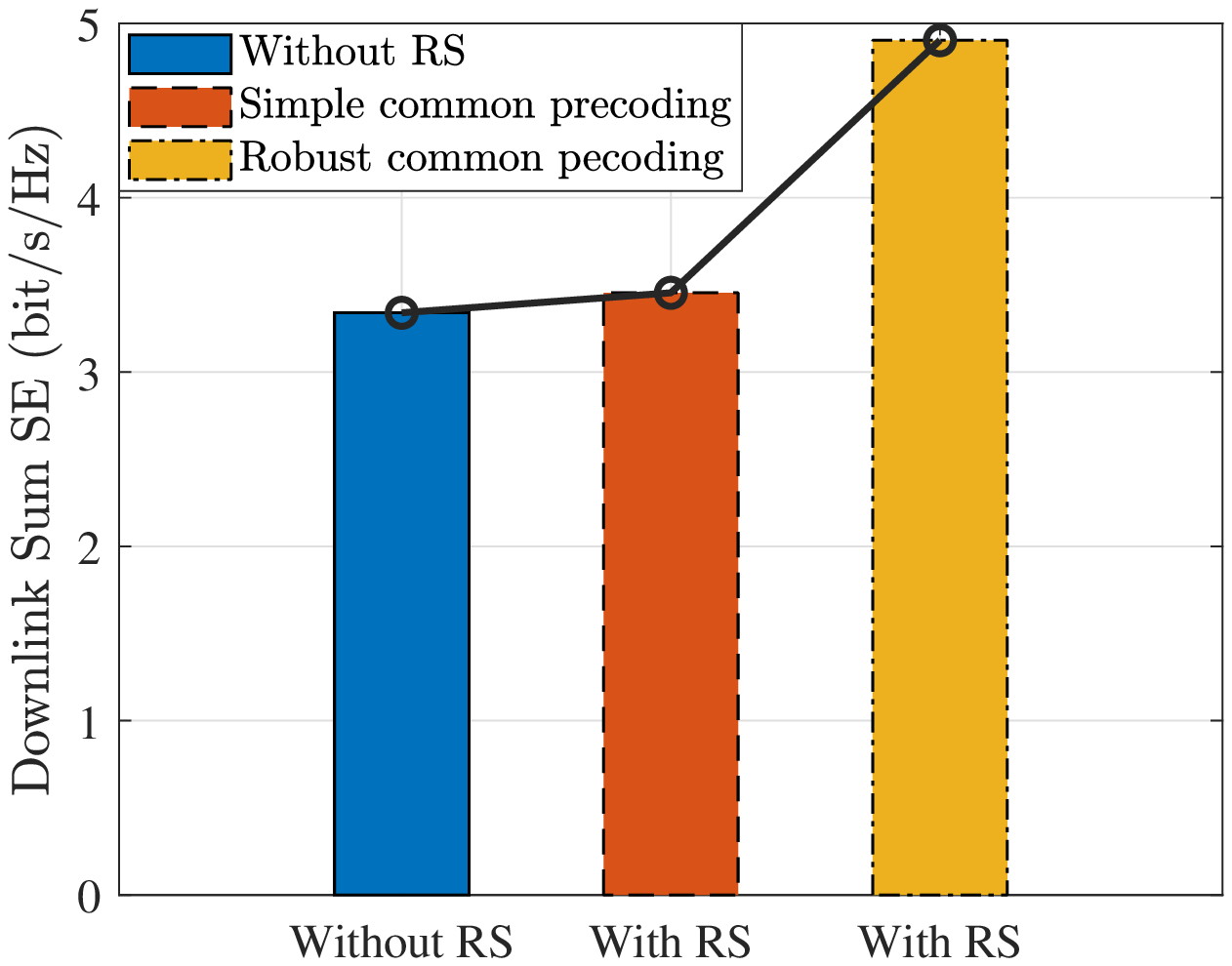}
\caption{Downlink sum SE for asynchronous CF massive MIMO systems with square setting for APs under coherent transmission ($L=40$, $K=8$, $N=2$, $\tau_p = 4$).} \vspace{-4mm}
\label{threeSchemes}
\end{minipage}
\end{figure}

Fig.~\ref{RandLine} presents the downlink sum SE of asynchronous CF massive MIMO with our proposed robust precoding for the  common message under the coherent transmission and DU-MR private precoding. In addition, the case without RS and the case with simple common precoding are considered for comparison. We first assume that the APs are randomly deployed in the square simulation area. It is found that the downlink sum SE of RS with the simple common precoding is increased by $1.5$ bit/s/Hz compared to that without RS. Yet, the downlink sum SE of RS with robust common precoding is only increased by $0.2$ bit/s/Hz compared to that with simple common precoding. The reason is that the SE of the common messages in RS is limited by the channel condition of the worst UE for ensuring common messages are successfully decoded by all the UEs. Fortunately, uniform coverage properties of CF massive MIMO systems make RS suitable for deployment with simple common precoding to obtain large performance gains. Besides, we consider an extreme scenario with a non-uniform coverage where the APs are deployed in a radio strip on the one side of the square simulation area. It is clear that the sum SE gain of the simple common precoding is less than half of the sum SE gain of the robust common precoding. Actually, the oscillator phase of different UEs is different, which also increases performance differences among UEs. From Fig.~\ref{RandLine}, it can be observed that when the oscillator of each UEs is different, the sum SE gain of the simple common precoding is only $0.1$ bit/s/Hz, whereas the sum SE gain of the robust common precoding can reach $1$ bit/s/Hz. Therefore, it is necessary to adopt robust common precoding in the case of serious oscillator mismatch.

The downlink sum SE of asynchronous CF massive MIMO with our proposed robust common precoding under coherent transmission and DF-MR private precoding is illustrated in Fig.~\ref{threeSchemes}. Due to the strict delay synchronization requirements imposed by the RS, the sum SE gain of the simple precoding for the common messages is not more than $0.1$ bit/s/Hz. However, the RS with the robust common precoding can effectively overcome the influence of delay asynchronous and obtain $1.6$ bit/s/Hz sum SE gain. Therefore, when the RS is deployed in CF massive MIMO systems with un-awared delay phase, robust precoding for the common messages is necessary.

\section{Conclusion}\label{se:conclusion}

We investigated the performance of CF massive MIMO systems with asynchronous reception, including both the delay and oscillator phases. Besides, the RS strategy relying on the transmission of common and private messages was adopted to reduce the multiuser interference caused by imperfect CSI. Moreover, a robust precoding for the common message was designed to mitigate the effect of asynchronous reception on CF massive MIMO systems.
Specifically, taking into account the coherent and non-coherent transmission, we first derived novel closed-form SE expressions for RS-assisted CF massive MIMO systems with channel estimation errors caused by phase-asynchronization and pilot contamination. It was shown that asynchronous reception destroys the pilot orthogonality and coherent data transmission resulting in poor system performance. In particular, obtaining an accurate delay phase is important for CF massive MIMO systems to realize coherent transmission. Moreover, it is interesting that the oscillator phase of UEs has a dominated effect on the SE performance than that of the APs while increasing the number of antennas can only significantly reduce the influence of the oscillator phase at the APs.
%Furthermore, in order to investigate the sum SE achieved by the RS strategy in the practical case with asynchronous reception, we derived the closed-form SE of both the common and private messages.
Furthermore, it was proved that RS significantly improves the performance of CF massive MIMO systems, but it is seriously affected by the asynchronous phases, especially the delay phase and the oscillator phase of the UEs.
Also, we designed a robust precoding to maximize the minimum individual SE of the common message.
It was found that the proposed robust precoding for the common messages performs well in some extreme scenarios, e.g., serious oscillator mismatch and unknown delay phase.
In our future work, we will investigate how the RS realizes scalable applications in user/cell-centric CF networks \cite{8761828}.
Besides, the impacts of imperfect SIC in RS systems should also be considered \cite{9932426}.

\begin{appendices}
\section{Proof of Theorem 1}

%Each UE is assumed to be aware of the channel statistics and the signal detection is performed with the required channel distribution information available.
We can derive the use-and-then-forget (UatF) capacity bound with ${\text{SINR}}_k^{\text{p}}\left[ n \right] $ is given by \cite{9322468}
\begin{align}
\left( {{p_{{\text{dp}}}}{{\left| {\sum\limits_{l = 1}^L {\underbrace {\mathbb{E}\!\left\{ {{\mathbf{g}}_{kl}^{\text{H}}\left[ n \right]\sqrt {{\mu _l}} {{\mathbf{v}}_{kl}}} \right\}}_{{\text{DS}}_{kl}^{\text{p}}\left[ n \right]}} } \right|}^2}} \right)\!/\!\left(\! {{p_{{\text{dp}}}}\sum\limits_{i = 1}^K {\underbrace {\mathbb{E}\!\left\{ {{{\left| {\sum\limits_{l = 1}^L {{\mathbf{g}}_{kl}^{\text{H}}\left[ n \right]\sqrt {{\mu _l}} {{\mathbf{v}}_{il}}} } \right|}^2}} \right\}}_{{\text{INT}}_i^{\text{p}}\left[ n \right]}}  \!-\! {p_{{\text{dp}}}}{{\left| {\sum\limits_{l = 1}^L {{\text{DS}}_{kl}^{\text{p}}\left[ n \right]} } \right|}^2} \!+\! \sigma _{\text{d}}^2} \!\right) , \notag
\end{align}
where ${{\text{D}}{{\text{S}}_{kl}^\text{p}}\left[ n \right]}$ denotes the desired signal and ${{\text{IN}}{{\text{T}}_i^\text{p}}\left[ n \right]}$ is the interference from other UEs\footnote{Note that $\text{DS}_k$ is a part of $\text{INT}_k$, which is subtracted by $\text{DS}_k$ to obtain the beamforming gain uncertainty as the part of residual interference part.}. It is assumed the DU-MR precoding ${{\mathbf{v}}_{kl}} = {\theta _{kl}}{{{\mathbf{\hat h}}}_{kl}}\left[ \lambda  \right]$ is used.
Then, the normalization parameter regarding the precoding in \eqref{mul} can be written as ${\mu _l} = {1}/{{\sum\limits_{i = 1}^K {{\text{tr}}\left( {{{\mathbf{Q}}_{il}}} \right)} }}$.
Substituting \eqref{gkl} into ${{\text{D}}{{\text{S}}_{kl}^\text{p}}\left[ n \right]}$, and using the definition and property of $\Theta _{kl}$ in \eqref{theta1} and \eqref{theta2}, we can derive
\begin{align}
   {{\text{D}}{{\text{S}}_{kl}^\text{p}}\left[ n \right]} &= \mathbb{E}\left\{ {{\mathbf{h}}_{kl}^{\text{H}}\left[ \lambda  \right]\Theta _{kl}^*\left[ n - \lambda \right]\theta _{kl}^*\sqrt {{\mu _l}} {\theta _{kl}}{{{\mathbf{\hat h}}}_{kl}}\left[ \lambda  \right]} \right\} \notag \\
   &= \mathbb{E}\left\{ {\Theta _{kl}^*\left[ n - \lambda \right]} \right\}\sqrt {{\mu _l}} \; \mathbb{E}\left\{ {{\mathbf{h}}_{kl}^{\text{H}}\left[ \lambda  \right]{{{\mathbf{\hat h}}}_{kl}}\left[ \lambda  \right]} \right\} = {e^{ - \frac{{n - \lambda }}{2}\left( {\sigma _{{\text{ap}}}^2 + \sigma _{{\text{ue}}}^2} \right)}}\sqrt {{\mu _l}} {\text{tr}}\left( {{{\mathbf{Q}}_{kl}}} \right) .
\end{align}
Moreover, with the help of \cite[Eq. (69)]{9322468}, we have
\begin{align}\label{INT}
  {\text{IN}}{{\text{T}}_i^\text{p}}\left[ n \right] &= \sum\limits_{l = 1}^L {\underbrace {\mathbb{E}\left\{ {{{\left| {{\mathbf{g}}_{kl}^{\text{H}}\left[ n \right]\sqrt {{\mu _l}} {\theta _{il}}{{{\mathbf{\hat h}}}_{il}}\left[ \lambda  \right]} \right|}^2}} \right\}}_{{\Upsilon _1}}} \notag\\
    &+ \sum\limits_{l = 1}^L {\mathop \sum \limits_{m \ne l}^L } \underbrace {\mathbb{E}\left\{ {{{\left( {{\mathbf{g}}_{kl}^{\text{H}}\left[ n \right]\sqrt {{\mu _l}} {\theta _{il}}{{{\mathbf{\hat h}}}_{il}}\left[ \lambda  \right]} \right)}^*}\left( {{\mathbf{g}}_{km}^{\text{H}}\left[ n \right]\sqrt {{\mu _m}} {\theta _{im}}{{{\mathbf{\hat h}}}_{im}}\left[ \lambda  \right]} \right)} \right\}}_{{\Upsilon _2}} .
\end{align}
Besides, using \eqref{gkl}, \eqref{theta1} and the property that $\left| {\exp \left( jx \right)} \right| = 1$ for any real number $x$ and $j$ being the imaginary unit, we derive
\begin{align}\label{gamma1}
  {\Upsilon _1} = \mathbb{E}\left\{ {{{\left| {{\mathbf{h}}_{kl}^{\text{H}}\left[ \lambda  \right]\Theta _{kl}^*\left[ n - \lambda \right]\theta _{kl}^*\sqrt {{\mu _l}} {\theta _{il}}{{{\mathbf{\hat h}}}_{il}}\left[ \lambda  \right]} \right|}^2}} \right\} = {\mu _l}\mathbb{E}\left\{ {{{\left| {{\mathbf{h}}_{kl}^{\text{H}}\left[ \lambda  \right]{{{\mathbf{\hat h}}}_{il}}\left[ \lambda  \right]} \right|}^2}} \right\} .
\end{align}
Following the similar steps of \cite[Eq. (62)]{9322468}, we can write \eqref{gamma1} as
\begin{align}\label{gamma11}
{\Upsilon _1} = {\mu _l}{\text{tr}}\left( {{{\mathbf{Q}}_{il}}{{\mathbf{R}}_{kl}}} \right) + \left\{ {\begin{array}{*{20}{c}}
  {{\mu _l}{{\left| {{\text{tr}}\left( {{{{\mathbf{\bar Q}}}_{kil}}} \right)} \right|}^2},i \in {\mathcal{P}_k}} \\
  {0,i \notin \mathcal{P}} .
\end{array}} \right.
\end{align}
By using \eqref{gkl} and \eqref{theta1}, we obtain
\begin{align}\label{gamma2}
  {\Upsilon _2} &= {\theta _{kl}}\theta _{il}^*\theta _{km}^*{\theta _{im}}\sqrt {{\mu _l}{\mu _m}} \;\underbrace {\mathbb{E}\left\{ {{\Theta _{kl}}\left[ n - \lambda \right]\Theta _{km}^*\left[ n - \lambda \right]} \right\}}_{{\Upsilon _3}} \notag\\
  & \times {\underbrace {\mathbb{E}\left\{ {{\mathbf{h}}_{kl}^{\text{H}}\left[ \lambda  \right]{{{\mathbf{\hat h}}}_{il}}\left[ \lambda  \right]} \right\}}_{{\Upsilon _4}}}^*\mathbb{E}\left\{ {{\mathbf{h}}_{km}^{\text{H}}\left[ \lambda  \right]{{{\mathbf{\hat h}}}_{im}}\left[ \lambda  \right]} \right\} .
\end{align}
By utilizing the definition and property of $\Theta_{kl}$ in \eqref{theta1} and \eqref{theta2}, we derive
\begin{align}\label{gamma3}
  {{\Upsilon _3}} &= \mathbb{E}\left\{ {{e^{j\sum\limits_{s = \lambda  + 1}^n {\left( {\delta _k^{{\text{ue}}}\left[ s \right] + \delta _l^{{\text{ap}}}\left[ s \right]} \right)} }}{e^{ - j\sum\limits_{s = \lambda  + 1}^n {\left( {\delta _k^{{\text{ue}}}\left[ s \right] + \delta _m^{{\text{ap}}}\left[ s \right]} \right)} }}} \right\} = \mathbb{E}\left\{ {{e^{j\sum\limits_{s = \lambda  + 1}^n {\left( {\delta _l^{{\text{ap}}}\left[ s \right] - \delta _m^{{\text{ap}}}\left[ s \right]} \right)} }}} \right\} \notag \\
   & = \mathbb{E}\left\{ {{e^{j\sum\limits_{s = \lambda  + 1}^n {\delta _l^{{\text{ap}}}\left[ s \right]} }}} \right\}\mathbb{E}\left\{ {{e^{ - j\sum\limits_{s = \lambda  + 1}^n {\delta _m^{{\text{ap}}}\left[ s \right]} }}} \right\} = {e^{ - \frac{{n - \lambda }}{2}\sigma _{{\text{ap}}}^2}}{e^{ - \frac{{n - \lambda }}{2}\sigma _{{\text{ap}}}^2}} = {e^{ - \left( {n - \lambda } \right)\sigma _{{\text{ap}}}^2}} .
\end{align}
Based on the properties of MMSE estimation with ${{{\mathbf{\hat h}}}_{kl}}\left[ \lambda  \right]$ and ${{{\mathbf{\tilde h}}}_{kl}}\left[ \lambda  \right]$ are independent, we have
\begin{align}\label{gamma4}
{\Upsilon _4} = \mathbb{E}\left\{ {{\mathbf{\hat h}}_{kl}^{\text{H}}\left[ \lambda  \right]{{{\mathbf{\hat h}}}_{il}}\left[ \lambda  \right]} \right\} = {\text{tr}}\left( {\mathbb{E}\left\{ {{{{\mathbf{\hat h}}}_{il}}\left[ \lambda  \right]{\mathbf{\hat h}}_{kl}^{\text{H}}\left[ \lambda  \right]} \right\}} \right) .
\end{align}
Substituting \eqref{hhat} into \eqref{gamma4}, we then obtain
\begin{align}\label{gamma44}
{\Upsilon _4} = \left\{ {\begin{array}{*{20}{c}}
  {{\theta _{kl}}\theta _{il}^{\text{*}}{\text{tr}}\left( {{{{\mathbf{\bar Q}}}_{kil}}} \right),i \in {\mathcal{P}_k}} \\
  {0,i \notin {\mathcal{P}_k}} .
\end{array}} \right.
\end{align}
Finally, with the help of \eqref{gamma3} and \eqref{gamma44} and plugging \eqref{gamma11} and \eqref{gamma2} into \eqref{INT}, we derive
\begin{align}\label{INTc}
&{\text{IN}}{{\text{T}}_i^\text{p}}\left[ n \right] = \sum\limits_{l = 1}^L {\mu _l}{\text{tr}}\left( {{{\mathbf{Q}}_{il}}{{\mathbf{R}}_{kl}}} \right) \notag\\
& + \left\{ {\begin{array}{*{20}{c}}
  {\left( {1 - {e^{ - \left( {n - \lambda } \right)\left( {\sigma _{{\text{ap}}}^2} \right)}}} \right)\sum\limits_{l = 1}^L {{\mu _l}{{\left| {{\text{tr}}\left( {{{{\mathbf{\bar Q}}}_{kil}}} \right)} \right|}^2}}  + {e^{ - \left( {n - \lambda } \right)\left( {\sigma _{{\text{ap}}}^2} \right)}}{{\left| {\sum\limits_{l = 1}^L {\sqrt {{\mu _l}} {\text{tr}}\left( {{{{\mathbf{\bar Q}}}_{kil}}} \right)} } \right|}^2},i \in {\mathcal{P}_k}}, \\
  {0,i \notin {\mathcal{P}_k}},
\end{array}} \right.
\end{align}
and this completes the proof.

\section{Proof of Theorem 3}

Using the UatF bound \cite{9322468}, we can derive the ${\text{SINR}}_k^{\text{c}}\left[ n \right]$ of the common messages as
\begin{align}
\left(\!\! {{p_{{\text{dc}}}}\!{{\left|\! {\sum\limits_{l = 1}^L {\underbrace {\!\mathbb{E}\!\left\{\! {{\mathbf{g}}_{kl}^{\text{H}}\!\left[ n \right]\!\!\sqrt {{\eta _l}} {{\mathbf{v}}_{{\text{c}},l}}} \!\right\}}_{{\text{DS}}_{kl}^{\text{c}}\left[ n \right]}} } \!\right|}^2}} \right)\!\!/\!\!\left(\!\! {{p_{{\text{dc}}}}\underbrace {\!\mathbb{E}\!\!\left\{\! {{{\left|\! {\sum\limits_{l = 1}^L \!{{\mathbf{g}}_{kl}^{\text{H}}\!\left[ n \right]\!\!\sqrt {{\eta _l}} {{\mathbf{v}}_{{\text{c}},l}}} } \right|}^2}} \!\right\}}_{{\text{INT}}_k^{\text{c}}\left[ n \!\right]} \!- {p_{{\text{dc}}}}\!{{\left|\! {\sum\limits_{l = 1}^L \!{{\text{DS}}_{kl}^{\text{c}}\!\left[ n \right]} } \!\right|}^2} \!\!+\! {p_{{\text{dp}}}}\!\sum\limits_{i = 1}^K \!{{\text{INT}}_i^{\text{p}}\!\left[ n \right] \!+\! \sigma _{\text{d}}^2} } \!\!\right) \!, \notag
\end{align}
where ${\text{DS}}_{kl}^{\text{c}}\left[ n \right]$ is the desired common signal. Besides, ${\text{INT}}_k^{\text{c}}\left[ n \right]$ is subtracted by ${\text{DS}}_{kl}^{\text{c}}\left[ n \right]$ to obtain the residual interference caused by beamforming gain uncertainty \cite{9322468}. Note that ${\text{IN}}{{\text{T}}_i^\text{p}}\left[ n \right]$ is the interference from the private signal, which is given by \eqref{INTc}. It is assumed that the DU-MR private precoding ${{\mathbf{v}}_{kl}} = {\theta _{kl}}{{{\mathbf{\hat h}}}_{kl}}\left[ \lambda  \right]$ and linear common precoding ${{\mathbf{v}}_{{\mathrm{c}},l}} = \sum\nolimits_{i = 1}^K {a_{il}{{\mathbf{v}}_{il}}}$ is used.
Substituting \eqref{gkl} into ${\text{DS}}_{kl}^{\text{c}}\left[ n \right]$ and with the help of \eqref{theta1}, \eqref{theta2}, and \eqref{gamma44}, we have
\begin{align}\label{DS_kl}
{\text{DS}}_{kl}^{\text{c}}\left[ n \right] = \sum\limits_{i = 1}^K {{a_{il}}\theta _{kl}^*{\theta _{il}}\mathbb{E}\left\{ {{\mathbf{h}}_{kl}^{\text{H}}\left[ \lambda  \right]{{{\mathbf{\hat h}}}_{il}}\left[ \lambda  \right]} \right\}\mathbb{E}\left\{ {\Theta _{kl}^*\left[ n \right]} \right\}}  = {e^{ - \frac{{n - \lambda }}{2}\left( {\sigma _{{\text{ap}}}^2 + \sigma _{{\text{ue}}}^2} \right)}}\sum\limits_{i \in {\mathcal{P}_k}}^K {{a_{il}}{\text{tr}}\left( {{{{\mathbf{\bar Q}}}_{kil}}} \right)} .
\end{align}
Moreover, we also can obtain the normalization parameter regarding the precoding in \eqref{etal} as
\begin{align}
 \eta_l = {1}/{\mathbb{E}\left\{ {{{\left\| {\sum\limits_{k = 1}^K {{a_{kl}}{\theta _{kl}}{{{\mathbf{\hat h}}}_{kl}}\left[ \lambda  \right]} } \right\|}^2}} \right\}} = {1}/{\sum\limits_{k = 1}^K {\sum\limits_{i \in {\mathcal{P}_k}}^K {{a_{kl}^{*}}{a_{il}}{\text{tr}}\left( {{{{\mathbf{\bar Q}}}_{kil}}} \right)} }} .
\end{align}
For deriving ${\text{INT}}_k^{\text{c}}\left[ n \right]$, we first expand it into
\begin{align}
{\text{INT}}_k^{\text{c}}\left[ n \right] = \sum\limits_{l = 1}^L {{\eta _l}\underbrace {\mathbb{E}\left\{ {{{\left| {{\mathbf{g}}_{kl}^{\text{H}}\left[ n \right]{{\mathbf{v}}_{{\text{c}},l}}} \right|}^2}} \right\}}_{{\Upsilon _5}}}  + \sum\limits_{l = 1}^L {\sum\limits_{m \ne l}^L {\sqrt {{\eta _l}{\eta _m}} \underbrace {\mathbb{E}\left\{ {{{\left( {{\mathbf{g}}_{kl}^{\text{H}}\left[ n \right]{{\mathbf{v}}_{{\text{c}},l}}} \right)}^*}\left( {{\mathbf{g}}_{km}^{\text{H}}\left[ n \right]{{\mathbf{v}}_{{\text{c}},m}}} \right)} \right\}}_{{\Upsilon _6}}} } .
\end{align}
Substituting \eqref{gkl} and ${{\mathbf{v}}_{{\text{c}},l}} = \sum\nolimits_{i = 1}^K {a_{il}{\theta _{il}}{{{\mathbf{\hat h}}}_{il}}\left[ \lambda  \right]} $ into $\Upsilon_5$, we have
\begin{align}
{\Upsilon _5} \!=\! \sum\limits_{i = 1}^K {{a_{il}}{a_{il}}\underbrace {\mathbb{E}\left\{ {{{\left| {{\mathbf{h}}_{kl}^{\text{H}}\left[ \lambda  \right]{{{\mathbf{\hat h}}}_{il}}\left[ \lambda  \right]} \right|}^2}} \right\}}_{{\Upsilon _7}}}  \!+\! \sum\limits_{i = 1}^K {\sum\limits_{j \ne i}^K {{a_{il}}{a_{jl}}\theta _{il}^*{\theta _{jl}}\underbrace {\mathbb{E}\left\{ {{{\left( {{\mathbf{h}}_{kl}^{\text{H}}\left[ \lambda  \right]{{{\mathbf{\hat h}}}_{il}}\left[ \lambda  \right]} \right)}^*}\left( {{\mathbf{h}}_{kl}^{\text{H}}\left[ \lambda  \right]{{{\mathbf{\hat h}}}_{jl}}\left[ \lambda  \right]} \right)} \right\}}_{{\Upsilon _8}}} } . \notag
\end{align}
With the help of \eqref{gamma3} and \eqref{DS_kl}, we obtain
\begin{align}
\Upsilon _6 = {e^{ - \left( {n - \lambda } \right)\sigma _{{\text{ap}}}^2}}\left( {\sum\limits_{i \in {\mathcal{P}_k}}^K {{a_{il}}{\text{tr}}\left( {{{{\mathbf{\bar Q}}}_{kil}}} \right)} } \right)\left( {\sum\limits_{j \in {\mathcal{P}_k}}^K {{a_{jm}}{\text{tr}}\left( {{{{\mathbf{\bar Q}}}_{kjm}}} \right)} } \right) .
\end{align}
Besides, $\Upsilon _7$ can be derived by \eqref{gamma11}. Following the similar steps, we can obtain $\Upsilon _8$ as
\begin{align}
\Upsilon _8 = {\theta _{il}}\theta _{jl}^*{\text{tr}}\left( {{{\mathbf{R}}_{kl}}{{{\mathbf{\bar Q}}}_{ijl}}} \right) + \left\{ {\begin{array}{*{20}{c}}
  {{\theta _{il}}\theta _{jl}^*{\text{tr}}\left( {{{{\mathbf{\bar Q}}}_{kil}}} \right){\text{tr}}\left( {{{{\mathbf{\bar Q}}}_{kjl}}} \right),i \in {\mathcal{P}_k}} \\
  {0,i \notin {\mathcal{P}_k}}
\end{array}} \right.,j \in {\mathcal{P}_i} .
\end{align}
Finally, we have
\begin{align}
{\text{INT}}_k^{\text{c}}\left[ n \right] &= \left( {1 - {e^{ - \left( {n - \lambda } \right)\sigma _{{\text{ap}}}^2}}} \right)\sum\limits_{l = 1}^L {{\eta _l}\sum\limits_{i \in {\mathcal{P}_k}}^K {\sum\limits_{j \in {\mathcal{P}_i}}^K {{a_{il}}{a_{jl}}{\text{tr}}\left( {{{{\mathbf{\bar Q}}}_{kil}}} \right){\text{tr}}\left( {{{{\mathbf{\bar Q}}}_{kjl}}} \right)} } }  \notag\\
&+ \sum\limits_{l = 1}^L {{\eta _l}\sum\limits_{i = 1}^K {\sum\limits_{j \in {\mathcal{P}_i}}^K {{a_{il}}{a_{jl}}{\text{tr}}\left( {{{\mathbf{R}}_{kl}}{{{\mathbf{\bar Q}}}_{ijl}}} \right)} } }  + {e^{ - \left( {n - \lambda } \right)\sigma _{{\text{ap}}}^2}}{\left| {\sum\limits_{l = 1}^L {\sqrt {{\eta _l}} \sum\limits_{i \in {\mathcal{P}_k}}^K {{a_{il}}{\text{tr}}\left( {{{{\mathbf{\bar Q}}}_{kil}}} \right)} } } \right|^2} ,
\end{align}
and the result follows immediately.
\end{appendices}

\vspace{0cm}
\bibliographystyle{IEEEtran}
\bibliography{IEEEabrv,Ref}

\end{document}